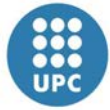 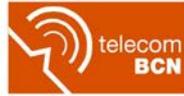

Escola Tècnica Superior d'Enginyeria
de Telecomunicació de Barcelona

UNIVERSITAT POLITÈCNICA DE CATALUNYA

# PROJECTE FINAL DE CARRERA

# Design and implementation of an Android application (MobilitApp+) to analyze the mobility patterns of citizens in the Metropolitan Region of Barcelona

*Estudis: Enginyeria de Telecomunicació*
*Autor:* Sergi Casanova Fouce
*Director/a:* Mónica Aguilar Igartua
*Any: Març 2015*

## RESUM DEL PROJECTE

En el nostre projecte hem dissenyat una aplicació per Android anomenada MobilitApp amb l'objectiu d'obtenir dades de mobilitat dels ciutadans de l'Àrea Metropolitana de Barcelona (AMB). La recollida de dades es realitza totalment en segon pla i s'obtenen, per una banda amb localitzacions periòdiques de l'usuari i, per l'altra, obtenint el tipus d'activitat que l'usuari està realitzant, de manera síncrona. Al final del dia, les dades són processades i finalment són enviades al servidor on s'emmagatzemaran en arxius per tal d'extreure'n posteriorment patrons de mobilitat que puguin ajudar a millorar les actuals infraestructures de transport.

L'aplicació és totalment estable i funcional tot i que la fiabilitat dels resultats obtinguts és millorable en alguns casos. En futures versions de l'aplicació es redefiniran els algoritmes utilitzant Machine Learning. D'aquesta manera, i mitjançant l'entrenament de dades, obtindrem una millora notable en els resultats.

## RESUMEN DEL PROYECTO

En nuestro proyecto hemos diseñado una aplicación para Android llamada MobilitApp con el objetivo de obtener datos de movilidad de los ciudadanos del Área Metropolitana de Barcelona (AMB). La recogida de los datos se realiza totalmente en segundo plano y se obtienen mediante localizaciones periódicas y la obtención del tipo de activad que realiza el usuario de manera síncrona. Al final del día, los datos son procesados y finalmente son enviados al servidor donde se almacenarán en archivos para posteriormente poder extraer patrones de movilidad que puedan ayudar a mejorar las actuales infraestructuras de transporte.

La aplicación es totalmente estable y funcional a pesar de que los resultados obtenidos son mejorables en algunos casos. En futuras versiones de la aplicación se redefinirán los algoritmos utilizando Machine Learning. De esta manera, y mediante el entreno de los datos, obtendremos una mejora notable en los resultados.

# ABSTRACT

In our project we have designed an Android application to obtain mobility data of the citizens in the metropolitan area of Barcelona. Our implementation synchronously obtains in background on the one hand, periodic location updates and, on the other hand, the type of activity citizens are doing. At the end of the day, all this data is processed and sent to a server where are stored to obtain mobility patterns from citizens that could help to improve the current transportation infrastructure.

MobilitApp is fully functional and stable although the results can be improved in some situations. In future releases we will implement machine learning technics to obtain significant improvements, especially in the activity recognition modules.

# Table of Contents

## LIST OF FIGURES

## LIST OF TABLES

## GLOSSARY OF ACRONYMS

**API** Application Programming Interface

**SDK** Software Development Kit

**APK** Android Application Package

**APP** Application

**ATM** Autoritat del Transport Metropolità

**AMB** Àrea Metropolitana de Barcelona

**GUI** Graphical User Interface

**GPS** Global Positioning System

**JSON** JavaScript Object Notation

**UPC** Universitat Politècnica de Catalunya

**Wi-Fi** Wireless Fidelity

**WPS** Wi-Fi-based Positioning System

**XML** eXtensible Markup Language

# Part I
# Introduction

## I.1.  Introduction

In this project, we are going to develop an Android application to obtain data regarding the mobility of citizens in Barcelona. In collaboration with ATM, this project's main objective is to use this data to determine mobility patterns that could be used to improve current transportation infrastructure.

Android Platform is used to develop MobilitApp for two main reasons:

- **Large Audience:** According to [0] in Spain at December 2014, 83% of the devices use Android.

- **Google APIs:** Google provides developers large amount of APIs (Application Programming Interface) to add different features to Apps. MobilitApp uses: Maps, Places, Directions, Location and Activity Recognition.

This study is part of the larger EMRISCO [1] project and will help ATM (Autoritat del Transport Metropolità) providing mobility data to improve the current transportation infrastructure. We added new features, updated algorithms, upgraded old interfaces, made visual enhancements and much more.
The main problem was to differentiate between motor vehicles or between railway transports. We had vehicle activity returned in car, motorbike or bus and transport activity returned in metro, tram or train.
In this continuation project, we have focused on improving these detection algorithms being able to differentiate between these activities.

These algorithms can always be improved with the aim of reducing detection errors. Using Google APis to detect activities could produce certain detection errors because these APIs make use of sensors and the GPS (Global Positioning System). It would be a good idea to radically upgrade all the algorithms using machine learning instead these APIs. Using machine learning is a difficult task and needs a huge amount of time and data to obtain the training set. Future releases of MobilitApp could include this improvement.

Once the detection algorithms are updated, we can proceed adding more features in order to make MobilitApp more attractive to users. Some of these features are: a History to be able to see on the map all the daily journeys, login and profile window and a graph where co2 saved and calories burned are displayed.

This document is structured with three main sections plus references and two annexes.
The first section is called MobilitApp main design criteria where we will explain some Android fundamentals such as fragments, services... and the four Google APIs used.
The second section is called MobilitApp architecture where all related with the application is explained with its respective code.
The third section will explain all the recognition algorithms for the different types of transportation: bus, metro, train and tram.
Finally we have two annexes, the first one will show five case studies, and the last one is the MobilitApp user guide.

# Part II
# MobilitApp Main Design

# II.1. Application Fundamentals

## II.1.1. Activity

**Description**

An activity [2] is an application component providing a view users can interact with. An application consists of multiple activities. The "main activity" is presented to the user when the application is launched for the first time. In our case, MobilitApp chooses between two "main" activities: Login Activity and the Main Activity. The first one is launched the first time MobilitApp is opened and the second one the following times. This Main Activity will start Location service and Activity Recognition service.

Each activity can start another activity in order to perform different actions. Each time a new activity starts, the previous activity is stopped, but the system preserves the activity in a stack (i.e. the "back stack"). When a new activity starts, it is pushed onto the back stack and takes user focus. When the user is done with the current activity and presses the Back button, it is popped from the stack and destroyed and the previous activity resumes.

Activities must be declared in the manifest file in order to be accessible to the system. This manifest file presents essential information about the app to the Android system. Information like: minimum SDK (Software Development Kit) version, permissions, activities, services…

**Lifecycle**

It is crucial to manage the lifecycle of our activities in order to develop a strong and flexible application. An activity has three states:

- **Resumed:** The activity is in the foreground of the screen and has user focus.

- **Paused:** Another activity is in the foreground (visible) and has focus, but this one is still visible. Paused activities are completely alive but can be killed by the system in low memory situations.

- **Stopped:** The activity is running in the background and is no longer visible by the user. Stopped activities are also alive and can be killed by the system.

In order to manage the lifecycle of our activity, we need to implement the callback methods. Callback methods can be overridden to do the appropriate work when the state of the activity changes. The most important callback methods are:

```java
public class ExampleActivity extends Activity {
    @Override
    public void onCreate(Bundle savedInstanceState) {
        super.onCreate(savedInstanceState);
        // The activity is being created.
    }
    @Override
    protected void onResume() {
        super.onResume();
        // The activity has become visible (it is now "resumed").
    }
    @Override
    protected void onPause() {
        super.onPause();
        // Another activity is taking focus (this activity is about to be "paused").
    }
    @Override
    protected void onDestroy() {
        super.onDestroy();
        // The activity is about to be destroyed.
    }
}
```

*onCreate()* method is called when the activity is created for the first time. This is where all the set up needs to be done such as create views, bind data to lists, and so on.

*onResume()* method is called just before the activity starts interacting with the user.

*onPause()* method is called when the system is about to start resuming another activity.

*onDestroy()* method is called before the activity is destroyed.

Figure 2.1 shows the activity lifecycle.

Figure 2.1: Activity Lifecycle

### II.1.2. Fragment

A Fragment [3] represents a portion of the user interface in an Activity. Multiple fragments can be combined in a single activity. Each fragment has its own lifecycle and can be added or removed while the activity is running.
MobilitApp uses fragments in all the activities because allows more dynamic and flexible designs.
There are two ways we can add a fragment to the activity layout:

- **Declaring the fragment inside the activity's layout file**

In this case, we can specify layout properties for the fragment as if it were a view. This is how we add the map on the activity.

```
<fragment
    android:id="@+id/map"
    android:layout_width="match_parent"
    android:layout_height="match_parent"
    class="com.google.android.gms.maps.MapFragment" />
```

- **Programmatically add the fragment to an existing ViewGroup**

At any time the activity is running, we can add fragments to our activity layout. We simply need to specify a ViewGroup in which to place the fragment.
Adding, replacing or removing actions are called fragments transactions and we need an instance of *FragmentTransaction* [4] to use methods such as: add, remove or replace.

```
getSupportFragmentManager().beginTransaction().add(R.id.containerhist, new
    HistoryFragment()).commit();
```

In this case, the *ViewGroup* is a *FrameLayout* declared in the activity's layout file and we use the "id" as a reference.

```
<FrameLayout
    android:id="@+id/containerhist"
    android:layout_width="match_parent"
    android:layout_height="wrap_content"
    android:layout_below="@id/toolbar"
    android:visibility="gone" >
</FrameLayout>
```

The fragment can access to its activity instance (and use activity methods) using *getActivity()*

**Lifecycle**

The most important callback methods are the same as with Activity plus one more:

*onCreateView()* method is called when it is time for the fragment to draw its user interface for the first time.

Figure 2.2: Fragment Lifecycle

### II.1.3. Service

A Service [5] is an application component that can perform long-running operations in the background and does not provide a user interface. A service can take two forms:

- **Started:** A service is started when an activity or a fragment starts it by calling *startService()* method. Once started, a service can run in the background indefinitely, even if the activity or fragment is destroyed. This is the form used by MobilitApp for Location Service and Activity Recognition Service.

- **Bound:** A service is "bound" when an activity or a fragment binds to it by calling *bindService()* method. A bound service offers a client-server interface that allows activities/fragments to interact with the service. This form is not used by MobilitApp.

There are two types of service: the normal (i.e. *Service*) and the *IntentService.*

The main difference between the two is that *Service* uses the application's main thread so it could slow the performance of any running activity. IntentService uses a worker thread to handle all start requests, one at a time. This work queue passes on intent at a time to our *onHandleIntent()* callback method.

**Lifecycle**

The most important callback methods are:

*onStartCommand()* method is called every time the service is started by calling *startService()*

*onHandleIntent()* method is called to process intents in the worker thread. When all requests have been handled stops itself.

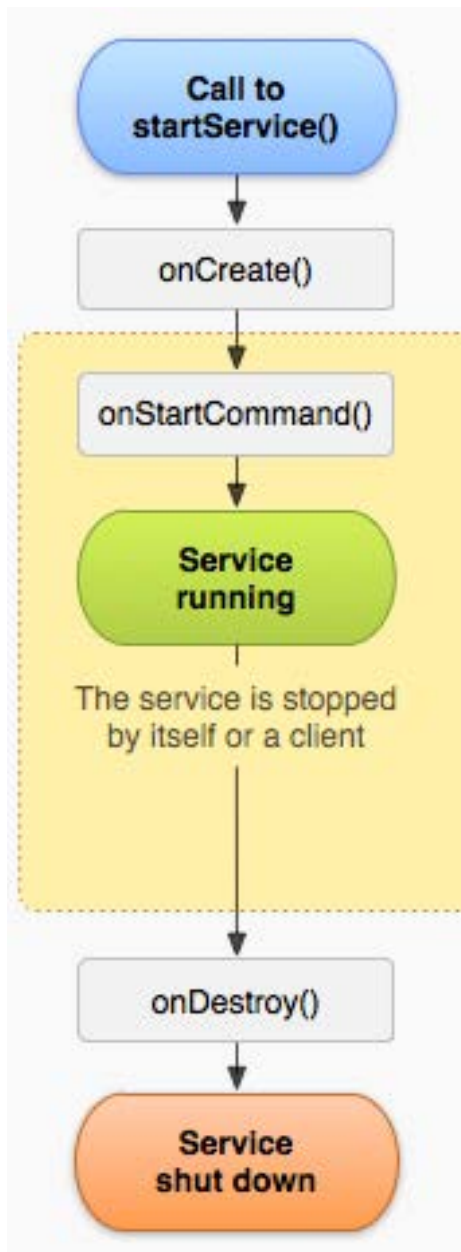

Figure 2.3: Service Lifecycle

### II.1.4. Localization

## Supporting different languages

It is always a good practice to present the application with more languages; thereby we can reach more users in Barcelona. MobilitApp [6] is available in three languages (i.e. Spanish, Catalan and English) depending on the language their mobile phone is set. It cannot be changed, so if the user's phone is in Catalan, MobilitApp dialog and text will be in Catalan too. If user's language is neither Spanish nor Catalan, the default language is set (i.e. English).

To make this possible [6] we need to create locale directories and string files. Within *res/* directory are subdirectories for various resource types (e.g. Layout resources are saved in *res/layout,* Drawable resources are saved in *res/drawable…).* String Resources define strings and string arrays and are saved in *res/values.* This is the default directory, so if we don't have any additional *values* directory this will be the one Android would use. An example of a bad practice is the following: all the strings in *values* directory are written in Spanish but user's mobile locale are set in English. In this situation user does not understand anything because MobilitApp dialogs and text are in Spanish and user only understands English. To solve this, we need to create additional *values* directories.

These additional *values* directories are followed by a hyphen and the ISO language code. If the language is spoken in different countries we add another hyphen and the region (preceded by lowercase "r". Figure 2.4 shows how Resources directory is structured in Android.

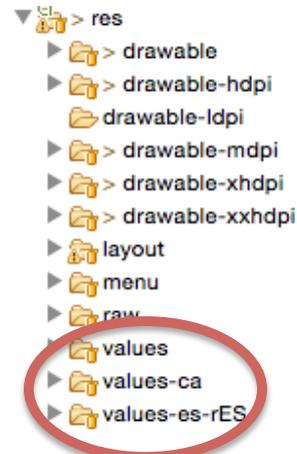

- **Spanish *values* directory:** values-es-rES
- **Catalan *values* directory:** values-ca

Figure 1.4: Resources Directory

Now we can create a *strings.xml* file in each directory with its respective language and then add string values for each locale into the appropriate file.
At runtime, the Android system uses the appropriate set of string resources based on the locale currently set for the user's device.

For example, the following are some different string resources for different languages.

```xml
<string name="acep">Aceptar</string>
<string name="cancel">Cancelar</string>
```

```xml
<string name="acep">Acceptar</string>
<string name="cancel">Cancel·lar</string>
```

```xml
<string name="acep">Accept</string>
<string name="cancel">Cancel</string>
```

Finally, we can reference our string resources in the code using the resource name defined by the string element's *name* attribute.

For example:

When we only need to supply the string resource to a method that requires a string.

```java
builder.setNegativeButton(R.string.cancel, null);
```

Or, when we need the string itself (i.e. the text) and not the resource.

```java
value[0] = getActivity().getResources().getString(R.string.calorie);
```

### II.1.5. Shared Preferences

Android provides several options to save persistent application data [7]. Choosing between these options depends on our specific needs, such as whether the data should be private or accessible to other applications and how much space our data requires.
Android provides 5 options to store data:

- **Shared Preferences:** Store private primitive data in key-value pairs.
- **Internal Storage:** Store private data on the device memory.
- **External Storage:** Store public data on the shared external storage.
- **SQLite Databases:** Store structured data in a private database.
- **Network Connection:** Store data on the web with your own network server.

MobilitApp use three of these five options:

- **Shared Preferences:** Store simple data such as profile information.
- **External Storage:** Store JSON (JavaScript Object Notation) files in *Downloads/* device directory.
- **Network Connection:** Store JSON files using Google Cloud Storage.

External Storage is explained in Segment section and Network Connection in Upload Section. So now, we need to explain the other store option used by MobilitApp: Shared Preferences.

The *SharedPreferences* class [8] provides a general framework that allows us to save and retrieve persistent key-value pairs of primitive data types. We can use *SharedPreferences* to save any primitive data: Booleans, floats, ints, longs and strings. This data will persist across user sessions, even if the application is closed or killed.

To get a *SharedPreferences* object for our applications we use the following method with two parameters: the name (i.e. filename) and the operating mode by default (i.e. MODE_PRIVATE).

```
prefsGoogle = getSharedPreferences("google_login", Context.MODE_PRIVATE);
```

If the files does not exist it will be created with the filename indicated.
In Private Mode the created file can only be accessed by the calling application (i.e. MobilitApp).

## Writing or Modification Values

1. Call *edit()* to get a *SharedPreferences.Editor* [7].

```
SharedPreferences.Editor editorG = prefsGoogle.edit();
```

2. Add values with methods such as *putBoolean()* or *putString()*. The first parameter is the key and the second is the value.

```
editorG.putString("birthday", p.getBirthday());
```

3. Commit the new values with *commit()*.

```
editorG.commit();
```

## Reading Values

1. We only need to use methods such as *getBoolean()* or *getString()*. The first parameter is the key and the second is the default value that will be returned when the key does not exist.

```
prefsGoogle.getString("nombre", "User");
```

## Output File Example

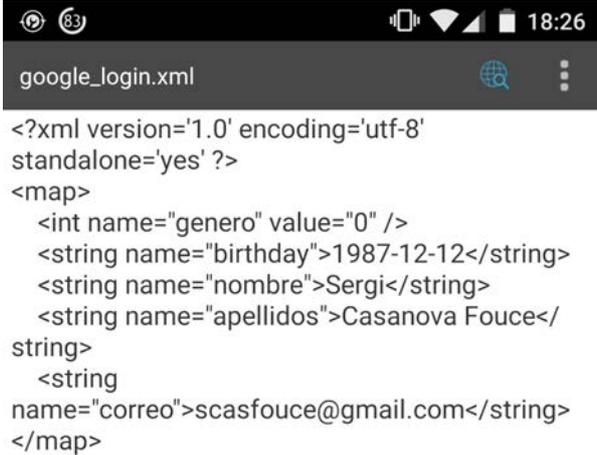

Figure 2.5: Shared Preferences Example

### II.1.6. Upload MobilitApp data to a Cloud Server

MobilitApp stores the data locally (in our smartphone) so, it mush implement a feature to upload data to a centralised storage server
This is particularly important if we want to make this data available for future analysis both at UPC (Universitat Politècnica de Catalunya) and at ATM. At the same time, in future releases we might want to provide the users with an historical of their journeys and some aggregated statistics.
Once this mobility data have been uploaded, ATM will be able to analyse and process it.
We are going to use Google Cloud Storage [9] as a first approximation of a real server. In the near future, MobilitApp data will be stored in a real server with its own web application to interact with the stored data.

All the information is uploaded anonymously even if the user is logged with Gmail or Facebook. The first time MobilitApp is installed generates a 256 bits code represented in hexadecimal (length 64) to be able to use it as an id. This will be the **user id** until the user uninstalls the App. In the Cloud Server each user has its own folder with his id as the folder name and inside this folder all the user data is stored daily.

Figure 2.6 shows an example of a folder with its name coded with SHA-256:

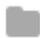

Figure 2.6: Google Cloud Storage folder

Inside this folder we have all the mobility files (one per day) stored in **JSON** format.
Figure 2.7 shows the contents of the above folder.

| NOMBRE | TAMAÑO | TIPO | ÚLTIMA SUBIDA |
|---|---|---|---|
| 16-12-2014.json | 2,55 KB | application/json | hace 10 días |
| 17-12-2014.json | 4,9 KB | application/json | hace 9 días |
| 18-12-2014.json | 5,12 KB | application/json | hace 8 días |
| 19-12-2014.json | 3,26 KB | application/json | hace 7 días |
| 20-12-2014.json | 15,39 KB | application/json | hace 6 días |
| 21-12-2014.json | 475 B | application/json | hace 5 días |
| 22-12-2014.json | 8,1 KB | application/json | hace 4 días |
| 23-12-2014.json | 6,85 KB | application/json | hace 3 días |
| 24-12-2014.json | 9,89 KB | application/json | hace 2 días |
| 25-12-2014.json | 1,57 KB | application/json | hace 1 día |

Figure 2.7: Folder Content

To be sure the collision probability is quite low (same user id for different users) when generating the user id with the 256-bit hash function the following method is implemented:

Our session id consists of the hash value of a random string. The random string is [10]:

- The current time (generated with the Java function *currentTimeMillis()*).
- A random number between 0 and 1 (generated with the Java function random).
- The process id number of the app (also basically a random number generated with the Java function myPid).
- A 256-bit key using an AES key generator.

And finally we create the hash of this concatenated string using **SHA-256**

```java
private String getSessionID() {

        String s;
        KeyGenerator keyGenerator = null;
        SecretKey key;

        // Generate a 256-bit key
        final int outputKeyLength = 256;
        SecureRandom secureRandom = new SecureRandom();
        try {
                keyGenerator = KeyGenerator.getInstance("AES");
        } catch (NoSuchAlgorithmException e) {
                e.printStackTrace();
        }
        keyGenerator.init(outputKeyLength, secureRandom);
        key = keyGenerator.generateKey();

        s = Long.toString(System.currentTimeMillis()) + Double.toString(Math.random())
                        + Integer.toString(Process.myPid()) + key.toString();
        return s;
}
```

*getSessionID()* method generates the random string using the four steps previously mentioned.

```java
private String hash(String s) {
        try {
                // Create SHA-256 Hash
                MessageDigest digest = java.security.MessageDigest.getInstance("SHA-256");
                digest.update(s.getBytes());
                byte messageDigest[] = digest.digest();
                // Create Hex String
                StringBuffer hexString = new StringBuffer();
                for (int i = 0; i < messageDigest.length; i++)
                        hexString.append(Integer.toHexString(0xFF & messageDigest[i]));
                return hexString.toString();
        } catch (NoSuchAlgorithmException e) {
                e.printStackTrace();
        }
        return "";
}
```

*hash()* method generates the user id with the SHA-256 hash.

## II.2. API

### II.2.1. Location Services API

#### II.2.1.1. *Location Sources in Android devices*

The selection of location sources is directly related to meet, on the one side, the low power consumption criterion, and on the other side, the specified requirements for mobility data. There are at least four location sources available in almost every Android device: GPS, WPS (Wi-Fi-based Positioning System), Cell ID and sensors. Each one of these sources has its own features regarding average power consumption, accuracy and coverage which basically depends on how location data is obtained as well as the minimum optimal update intervals to achieve with an acceptable level in quality of data.

Unfortunately, Android platform is not well documented regarding detailed profiles of average power consumption of each one of the location sources. For that reason, we are going to use approximations of their power usage profiles in order to select those that could meet our low power consumption criterion.

### Global Positioning System

GPS consists of up to 24 or more satellites broadcasting radio signals providing their locations, status and timestamp. The GPS receiver in the Android device can calculate the time difference between broadcast time and the time radio signal is received. When the device knows its distance from at least four different satellite signals, it can calculate its geographical position.

- **Average Power consumption:** GPS receiver works as an independent-powered component in the Android platform with the unique objective to obtain location samples. Therefore, all the battery power consumed by GPS is consequence of a user positioning operation. According to several references, we can estimate that the average power consumption of an active GPS with a location update interval between 5 and 15 seconds is approximately 125 to 145 mA.

- **Coverage:** GPS location coverage is probably the most limited of all four available sources. Although is a source that cover the whole globe, it only provides good performance in outdoor scenarios, because it requires satellites' visibility. GPS does not work in indoor locations (may provide some location data in very few of them if it has some kind of visibility of satellites, but with very poor performance) like buildings or undergrounds. Even in outdoor environments, GPS may suffer from poor performance, especially in cities, because objects like buildings, trees and other obstacles can overshadow visible satellites, decreasing its performance. Therefore, GPS range of coverage can be quite limited in a metropolitan region, compared to other sources, specifically if we want a continuous position tracking of the citizens.

- **Accuracy:** In outdoor scenarios with good satellite visibility, GPS is the positioning system that can provide the most accurate location of all four existing sources. In perfect conditions GPS can achieve a precision higher than 3 meters, but in optimal conditions the range is between 3 and 10 meters. In non-optimal conditions, i.e. lower visibility on the satellites, GPS accuracy range is typically between 30 and 100 meters. GPS is also capable to provide an estimation of altitude and speed along with location.

## Wi-Fi-based Positioning System

WPS sends a location request to Google location server with a list of MAC that are currently visible by the device, then the server compares this list with a list of known MAC addresses of the device itself, and identifies associated geocoded locations. After that, Google server uses these locations to triangulate the approximate location of the user.

- **Average Power consumption:** Wi-Fi receiver works as an independent-powered component in the Android platform but multiple-purpose that is not restricted only to obtain location samples, as happens with GPS receiver. Average power consumption of WPS represents only a small part of power usage of Wi-Fi services. According to multiple references, we are going to estimate that the average power consumption of an active WPS with a location update interval around 20 seconds is approximately 25 mA.

- **Coverage:** WPS location coverage is less broad than GPS but is more versatile. Its coverage it is not global but block level, because it has to be in the range of a wireless signal to be able to provide location updates. However, within this range, it can provide good performance both outdoor and indoor scenarios. Therefore, WPS range of coverage can be very useful when tracking continuous locations within cities and towns.

- **Accuracy:** In indoor scenarios, WPS is the positioning system that can provide the most accurate location of all four existing sources. In normal conditions, WPS can achieve a precision between 3 and 10 meters. In outdoors, the accuracy of GPS decreases significantly being able to provide a range between 25 and 150 meters, depending on the external conditions of the environment and the signal strength.

## Cell-ID based localization

Cell-ID based localization uses both the Location Area Code and Cell-ID that the Base Transceiver Station broadcasts. Android devices are always receiving these broadcast messages, which means that they always know their Cell-ID at any time. Knowing this, it can obtain an approximation of its actual location using the geographical coordinates of the corresponding Base Station.

- **Average Power consumption:** Since Android devices always know their Cell-ID; the average power consumption consequence of this positioning system is extremely low.

- **Coverage:** Cell-ID location coverage is global, as happens with GPS, and versatile, as happens with WPS, which implies that has the best coverage of all four existing location sources. Therefore, cell-ID has an adequate coverage range to continuously track location of citizens in the metropolitan region.

- **Accuracy:** Cell-ID based location system accuracy depends on cells size in the network that are connected with user's Android device. This means that in those regions with smaller cells (typically in cities and metropolitan regions), cell-ID location will provide good performance achieving an accuracy range between 50 and 200 meters. Moreover, in regions with bigger cells (typically in rural areas) performance will significantly decrease, and cell-ID will only achieve an accuracy range between 1000 meters and several kilometers.

### Sensors

- **Average Power consumption:** Android devices have usually several built-in sensors. However, not all of them can be used for positioning processes, being only accelerometer, magnetic field, orientation, and gyroscope that one that can be used for such purpose. Based on their specifications, the power consumption of these sensors are the following:

    - **Accelerometer** - 0.23 mA
    - **Magnetic Field** - 6.8 mA
    - **Orientation** - 13.13 mA
    - **Gyroscope** - 6.1 mA

    If all sensors were used together at the same time to obtain locations updates, we would have an average power consumption of 26.26 mA.

- **Coverage:** Sensors have a wide coverage because they can process multiple different types of data. From acceleration, that covers an individual range, to magnetic field, that covers a global range, sensors can provide data to improve location samples in almost any scenario.

- **Accuracy:** It depends on the type of sensor and how is used, which means it is infeasible to provide real accuracy values. However, according to Google, sensors improve the accuracy obtained with just simple WPS and Cell-ID.

#### II.2.1.2.    Android Location Tools

#### Fused Location Provider

This API considers all hardware components as single location sources layer that can be used according to predetermined criteria implemented on it. It offers four main features:

- **Simple API:** We can specify high-level needs like "high accuracy" or "low power", instead of having to worry about location providers.
- **Immediately available:** Gives immediate access to the best and most recent location.
- **Power-efficiency:** Minimizes app's use of power. Based on all incoming location requests and available sensors, fused location provider chooses the most efficient way to meet those needs.
- **Versatility:** Meets a wide range of needs, from foreground uses that need highly accurate location to background uses that need periodic location updates with negligible power impact.

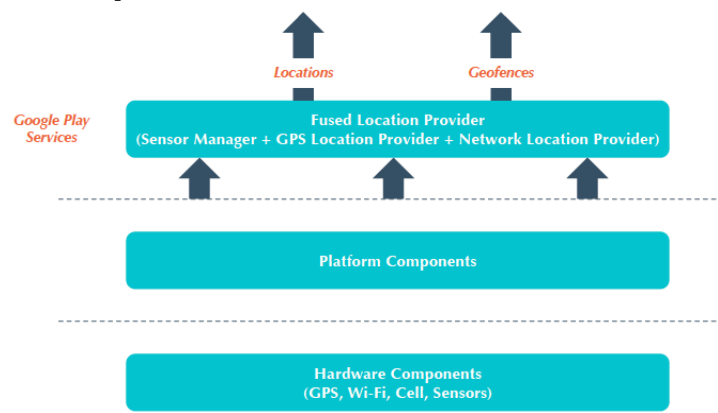

Figure 2.8: Fused Location Provider

**Location Request**

Location Request [11] is a data object that contains quality of service parameters for requests to the Fused Google Provider. The main parameters are:

- **Desired interval time:** This parameter establishes the ideal interval time between location updates. However, it is not an exact interval, which means that locations can be obtained in lower or higher interval times, depending on external conditions. According to our design criteria we should set this parameter to 20 seconds.

- **Fastest interval time:** This parameter establishes the minimum interval time between location updates. Unlike desired interval time, this is an exact interval, which means that location updates will not be obtained in lower intervals of time than set by this parameter. According to our design criteria we should also set this parameter to 20 seconds.

- **Priority mode:** This parameter establishes the priority of the location requests based on the location sources that we want to use and the accuracy of location samples we require. There are the following modes:

  o **High Accuracy:** This is the highest priority mode. Using this mode Location Client will obtain the best location available using all the location sources available together. This is also the mode with higher average power consumption because it is the only one that actually uses the GPS.

  o **Balanced Power Accuracy:** This mode has lower priority than High Accuracy mode. This mode only uses WPS, cell-ID and sensors to obtain location updates with a "block" level accuracy while keeping average power consumption very low.

  o **Low Power:** This mode has a lower priority than Balanced Power Accuracy mode. This mode only uses WPS, cell-ID and sensors to obtain location updates with a "city" level accuracy (10km) while keeping average power consumption extremely low.

  o **No Power:** This is the lowest priority mode. This mode does not use any location source to obtain location updates. Instead, obtains location updates requested by other third party applications in the Android platform. In this way the average power consumption is the lowest possible.

| PRIORITY | INTERVAL (s) | BATTERY DRAIN PER HOUR (%) | ACCURACY (m) |
|---|---|---|---|
| HIGH_ACCURACY | 5 | 7.25 | 10 |
| BALANCED_POWER_ACCURACY | 20 | 0.6 | 100 |
| LOW_POWER | - | small | 10.000 |
| NO_POWER | N/A | small | variable |

Table 2.1: Location Request priority modes

## II.2.2. Activity Recognition API

### II.2.2.1. Activity Recognition Sources in Android devices

In opposition of what happens with location, there is basically one main source to detect user's activity type: device sensors. Therefore, there is no real alternative to this source, and currently is the only possible option if we want to recognize activities. As we explained, there are four major sensors that are used to improve location updates, and although all of them can be somehow used for this purpose, in reality the main sensor that we are going to use is the accelerometer.

Accelerometer has the lowest power consumption of all sensors, but is also the one that requires more processing of samples to provide accurate results.

### II.2.2.2. Activity Recognition's Tools

Android platform provide developers with multiple tools, including services, libraries, classes and methods to implement and develop their application. However, these tools are limited to obtain raw sensor data. Process raw sensor data in order to detect user's activity with good accuracy is too complex to be designed and implemented to this project at the time of writing. Instead of that, we are going to use Google Activity Recognition API. These API provides a low power mechanism that can provide periodic updates of detected activity types by using low power sensors (basically accelerometer). We are going to process these samples in order to increase accuracy and efficiency. The most important classes that we are going to use from this API are presented below.

### Activity Recognition Result

This class [12] returns a list with all probable activities. It also provides a method that directly returns the most probable activity along with its confidence level called *getMostProbableActivity()*.

### Detected Activity

This class [13] returns an integer that indicates which activity type has bee detected by Activity Recognition API.

## II.2.3. Google Places API

The Google Places API [14] is a service that returns information about Places using HTTP requests. This API is the one used to find the nearest metro station so our desired Place is the metro.

HTTP response can be either in JSON format or in XML (eXtensible Markup Language) format, but Google recommends JSON format.

Unlike Google Directions API, this API uses an API key to identify our application. This is due to the fact that this API has a limit amount of requests. The quota is 100.000 requests per 24 hours period.

### II.2.3.1. Places Requests

First, we need to construct the URL using the parameters provided by Google.

Three parameters are required while others are optional; these parameters are separated using the ampersand character.

These three required parameters are:

- **key:** The key used to identify our application for purposes of quota management.

- **location:** Latitude/longitude around which to retrieve place information.
- **radius:** Defines the distance (in meters) within which to return place results.

MobilitApp also uses one optional parameter:

- **types:** Restricts the results to places matching at least one of the specified types.
  - **types=subway_station:** Used to search for metro stations.

```java
private String makeUrl(double latitude, double longitude, String place) {
        StringBuilder urlString = new StringBuilder(
                "https://maps.googleapis.com/maps/api/place/search/json?");

        urlString.append("&location=");
        urlString.append(Double.toString(latitude));
        urlString.append(",");
        urlString.append(Double.toString(longitude));
        urlStrig.append("&radius=150");
        urlString.append("&types=" + place);
        urlString.append("&key=" + API_KEY);

        return urlString.toString();
}
```

Now, we need to connect to this URL using *URLConnection* class. This class will be explained in the next chapter: Directions API.

## II.2.3.2. Places Response

If there is a metro station within a radius of 150 meters, we will receive a JSON format response. This response contains several fields but we need only one: the name of the metro station.
When there are two or more metro stations within this radius our algorithm only gets the name of the first one (i.e. the nearest one).

The following response will show the nearest metro station within 150 meters given one location

```
{
"html_attributions" : [],
"results" : [
        {
          "geometry" : {
                 "location" : {
                          "lat" : 41.384434,
                          "lng" : 2.11167
                 }
          },
          "icon" : "http://maps.gstatic.com/mapfiles/place_api/icons/generic_business-71.png",
          "id" : "377f10e5b229e754b2ed5cdda54ccf8394fe1bad",
          "name" : "Zona Universitària",
          "photos" : [
                 {
                   "height" : 2048,
                   "html_attributions" : [ "De un usuario de Google" ],
                   "photo_reference" : " ",
                   "width" : 1536
                 }
               ],
          "place_id" : "ChIJwYRtilaYpBIRgr0jGR0kkIk",
          "reference" : " ",
          "scope" : "GOOGLE",
          "types" : [ "subway_station", "train_station", "transit_station", "establishment" ],
          "vicinity" : "Spain"
          }
     ],
     "status" : "OK"
}
```

### II.2.3.3. Find Places Algorithm

Given two locations (origin and destination), this algorithm will return a dimension two *ArrayList* if both origin and destination responses have "OK" station (i.e. its nearest metro station is found). If only one response is "OK" the returned *ArrayList* will be dimension one and finally if neither origin nor destination responses have "OK" status null array is retuned.

```java
public ArrayList<String> findPlaces(double latitude, double longitude, double latitude2, double
                longitude2, String placeSpacification) {
        String urlString = makeUrl(latitude, longitude, placeSpacification);
        String urlString2 = makeUrl(latitude2, longitude2, placeSpacification);
        ArrayList<String> arrayList = new ArrayList<String>();
        try {
                JSONObject object = new JSONObject(getJSON(urlString));
                JSONArray array = object.getJSONArray("results");
                JSONObject object2 = new JSONObject(getJSON(urlString2));
                JSONArray array2 = object2.getJSONArray("results");
                try {
                        Place place = Place.jsonToPlace((JSONObject) array.get(0));
                        Place place2 = Place.jsonToPlace((JSONObject) array2.get(0));
                        if (place != null) { arrayList.add(place.getName()); }
                        if (place2 != null) { arrayList.add(place2.getName()); }
                } catch (Exception e) {
                }
                return arrayList;
        } catch (JSONException e) {
        }
        return null;
}
```

*jsonToPlace()* static method loops through the JSON response and returns the field with the name of the metro station.

```java
public static Place jsonToPlace(JSONObject point) {
        try {
                Place result = new Place();
                JSONObject geometry = (JSONObject) point.get("geometry");
                JSONObject location = (JSONObject) geometry.get("location");
                result.setLatitude((Double) location.get("lat"));
                result.setLongitude((Double) location.get("lng"));
                result.setName(point.getString("name"));
                return result;
        } catch (JSONException e) {
                e.printStackTrace();
        }
        return null;
}
```

## II.2.4. Google Directions API

The Google Directions API [15] is a service that calculates directions between location points using an HTTP request. MobilitApp uses this API to search for directions for two modes of transportations (i.e. Bus and Tram) and can return multiple routes between origin and destination.

HTTP response can be either in JSON format or in XML format, but Google recommends JSON format.

### II.2.4.1. Directions Requests

First, we need to construct the URL using the parameters provided by Google.

Two parameters are required while others are optional; these parameters are separated using the ampersand character.

These two required parameters are:

- **origin:** Latitude/longitude value from which we wish to calculate directions.

- **destination:** Latitude/longitude value to which we wish to calculate directions.

MobilitApp also uses other optional parameters:

- **departure_time:** Specifies the desired time of departure.
- **mode:** Specifies the mode of transport to use when calculating directions.
    - **mode=transit:** Requests directions via public transit routes.
- **alternatives:** If set to true, specifies that the Directions service may provide more than one route alternative
    - **alternatives=true**

```java
private String makeUrl(double latitude, double longitude, double latitude2, double
        longitude2, long departure_time) {

    StringBuilder urlString = new
            StringBuilder("https://maps.googleapis.com/maps/api/directions/json?");

    urlString.append("&origin=");
    urlString.append(Double.toString(latitude));
    urlString.append(",");
    urlString.append(Double.toString(longitude));
    urlString.append("&destination=");
    urlString.append(Double.toString(latitude2));
    urlString.append(",");
    urlString.append(Double.toString(longitude2));
    urlString.append("&mode=transit");
    urlString.append("&departure_time=");
    urlString.append(Long.toString(departure_time));
    urlString.append("&alternatives=true");

    return urlString.toString();
}
```

Now that we have constructed the URL it is time to make the connection. For this purpose exists the *URLConnection* class and to receive the response a buffer is implemented.

The following code will show the connection:

```java
private String getUrlContents(String theUrl) {
        StringBuilder content = new StringBuilder();
        try {
                URL url = new URL(theUrl);
                URLConnection urlConnection = url.openConnection();
                BufferedReader bufferedReader = new BufferedReader(new
                        InputStreamReader(urlConnection.getInputStream()));
                String line;
                while ((line = bufferedReader.readLine()) != null) {
                        content.append(line + "\n");
                }
                bufferedReader.close();
        } catch (Exception e) {
                e.printStackTrace();
        }
        return content.toString();
}
```

### II.2.4.2. Directions Response

JSON responses are very long so, we are going to explain the response fields used by MobilitApp with the definition and at the end a piece of the JSON response of the transit_details field.

- **routes:** Contains an array of routes from the origin to the destination.

Each route within the routes field may contain the following important field:

- **legs[]:** Contains different ways to make the journey from the origin to the destination.

Each leg may contain the following important field:

- **steps:** Contains an array of steps denoting information about each separate step of the leg.

The important field in the steps array is the following:

- **transit_details:** Contains transit specific information and it is only returned when travel mode is set to transit.

Here we have two important fields:

- **departure_time:** Contains the departure time for this leg of the journey.
- **line:** Contains information about the transit line used in this step.

Line field contains the most important field for us:

- **vehicle:** Is the array that contains the field **type**, this file contains the type of vehicle used on this line.

The algorithm uses only these two fields: departure_time and type. Our mission is to search for these two parameters in the *ArrayList* [16] and check if Bus or Tram is written or not as long as our departure time (time of the first location in a vehicle sample) is next to departure_time field.

```
"transit_details" : {
        "arrival_stop" : {
                "location" : {
                        "lat" : 41.4306046,
                        "lng" : 2.1449129
                },
                "name" : "Montbau"
        },
        "arrival_time" : {
                "text" : "21:52",
                "time_zone" : "Europe/Madrid",
                "value" : 1423774348
        },
        "departure_stop" : {
                "location" : {
                        "lat" : 41.4417677,
                        "lng" : 2.1658515
                },
                "name" : "Canyelles"
        },
        "departure_time" : {
                "text" : "21:47",
                "time_zone" : "Europe/Madrid",
                "value" : 1423774067
        },
        "headsign" : "Zona Universitària",
        "headway" : 420,
        "line" : {
                "agencies" : [
                        {
                                "name" : "TMB",
                                "phone" : "011 34 902 07 50 27",
                                "url" : "http://www.tmb.cat/"
                        }
                ],
                "color" : "#1eb53a",
                "name" : "Zona Universitària - Trinitat Nova",
                "short_name" : "L3",
                "text_color" : "#ffffff",
                "url" : "http://www.tmb.cat/ca/detall-linia-metro/-/linia/L3",
                "vehicle" : {
                        "icon" : "//maps.gstatic.com/mapfiles/transit/iw/6/subway.png",
                        "local_icon" : "//maps.gstatic.com/mapfiles/transit/iw/6/es-metro.png",
                        "name" : "Metro",
                        "type" : "SUBWAY"
                }
        },
        "num_stops" : 3
}
```

### II.2.4.3. Find Bus or Tram Algorithm

```java
public String findBusorTram() {
        String urlString = makeUrl(lat, lon, lat2, lon2, departure_time);
        long jsondeparture, resta1, jsonarrival, resta2;
        String type = "none";
        try {
                String json = getJSON(urlString);
                JSONObject object = new JSONObject(json);
                JSONArray array = object.getJSONArray("routes");
                for (int i = 0; i < array.length(); i++) {
                        JSONObject obj = array.getJSONObject(i);
                        JSONArray legs = obj.getJSONArray("legs");
                        JSONArray steps = legs.getJSONObject(0).getJSONArray("steps");

                        for (int j = 0; j < steps.length(); j++) {
                                if (steps.getJSONObject(j).has("transit_details")) {
                                        type =  steps.getJSONObject(j)
                                                .getJSONObject("transit_details")
                                        .getJSONObject("line")
                                                .getJSONObject("vehicle")
                                                .getString("type");

                                        if (type.equalsIgnoreCase("BUS") ||
                                                type.equalsIgnoreCase("TRAM")) {
                                                jsondeparture = steps.getJSONObject(j)
                                                        .getJSONObject("transit_details")
                                                        .getJSONObject("departure_time")
                                                        .getLong("value");
                                                jsonarrival = steps.getJSONObject(j)
                                                        .getJSONObject("transit_details")
                                                        .getJSONObject("arrival_time")
                                                        .getLong("value");

                                                resta1 = jsondeparture - departure_time;
                                                resta2 = jsonarrival - arrival_time;
                                                if (resta1 < 0)
                                                        resta1 = -resta1;
                                                if (resta2 < 0)
                                                        resta2 = -resta2;
                                                // error margin= 5 min at departure
                                                // error margin= 3 min at arrival
                                                if (resta1 < (5 * 60) && resta2 < (3 * 60)))
                                                        return type;
                                }       }       }       }
        } catch (JSONException e) { e.printStackTrace(); }
        return "none";
}
```

The algorithm returns a variable string depending on the field **type**, if "BUS" or "TRAM" does not appear in this field, "none" is returned.

We have an error margin with the departure time because the bus/tram stops at the bus/tram stop and then accelerates slowly. In this situation the detected activity could be still or on_foot and not the desired one: vehicle. Giving an error margin of 5 minutes for departure and 3 minutes for arrival assures we do not take into account these wrong samples.

# Part III
# MobilitApp Architecture

## III.1. Location Service

In this chapter we are going to explain in detail the most important part of our MobilitApp application, the location service. It's important because all the recognition algorithms are implemented here as well as the upload part and the post processing.

### III.1.1. Starting Location Service

Location service is initiated the first time we open MobilitApp by instantiating the class Requester and using the method *requestUpdates(...)* in our MainActivity fragment.
Requester class is instantiated inside the method *onActivityCreated(...)*

```
mrequesterLOC = new Requester(getActivity());
```

*getActivity()* is used in order to get the Context inside a fragment.

Once we have checked Google Play Services are installed in the device we can proceed to request location updates with:

```
mrequesterLOC.requestUpdates("LOC", false);
```

With these parameters we request location updates and GPS is set to false.

Requester is used for both Location updates and Activity Recognition updates depending on the first parameter in *requestUpdates()*. If "LOC" is written Location updates are initialized and Activity updates if "AR" is written.

To be able to get Location Updates a Google API Client is needed and then request a connection to it.
This client is generated by instantiating the builder of the *GoogleApiClient* class [17] adding the two APIs we are going to use and registering the two mandatory callbacks:

```
myClient = new GoogleApiClient.Builder(mContext).addApi(ActivityRecognition.API)
        .addApi(LocationServices.API).addConnectionCallbacks(this)
        .addOnConnectionFailedListener(this).build();
```

Once the client [18] is created we need to call the *connect()* method so, when the connection is done a call-back is generated in our *onConnected()* method. Here we can set the updates and finally disconnect the client.

```
if (type.equalsIgnoreCase("AR")) {
        ActivityRecognition.ActivityRecognitionApi.requestActivityUpdates(getClient(),AppUtils.
        UPDATE_INTERVAL_ACTIVITY, createRequestPendingIntent(type));
} else {
        LocationServices.FusedLocationApi.requestLocationUpdates(getClient(),
        createLR(mgps),createRequestPendingIntent(type));
        }
        // Disconnect the client
        requestDisconnection();
```

The best way requesting updates in background is using *PendingIntent* [19]. This *PendingIntent* is created in *createRequestPendingIntent()* method. This method has only one parameter in order to differentiate between the two APIs.

```java
private PendingIntent createRequestPendingIntent(String type) {
        PendingIntent pendingIntent;
                // If the PendingIntent already exists
        if (null != getRequestPendingIntent()) {
                // Return the existing intent
                return mActivityRecognitionPendingIntent;
                // If no PendingIntent exists
        } else {
                if (type.equalsIgnoreCase("AR")) {
                        Intent intent = new Intent(mContext, ActivityRecognitionService.class);
                        pendingIntent = PendingIntent.getService(mContext, 1, intent,
                                PendingIntent.FLAG_UPDATE_CURRENT);
                        setRequestPendingIntent(pendingIntent);
                } else {

                        Intent intent = new Intent(mContext, LocationService.class);
                        pendingIntent = PendingIntent.getService(mContext, 0, intent,
                                PendingIntent.FLAG_UPDATE_CURRENT);
                        setRequestPendingIntent(pendingIntent);
                }
                return pendingIntent;
        }
}
```

First, we need to create an intent to one of our two Services (Activity Recognition or Location) and then pass that intent to the pending intent generated from the static method *getService()*.

So, to summarize, every time we call requestUpdates PendingIntents are generated to its service.

But, How frequently are these PendingIntents generated?

In Location ones there are two possible intervals according to our design criteria, the normal interval: 20 seconds and the GPS interval: 5 seconds.

We are going to configure these intervals in *createLR()* method:

```java
private LocationRequest createLR(boolean gps) {
        LocationRequest locationRequest = null;
        if (!gps) {
                locationRequest =LocationRequest.create()
                                .setPriority(LocationRequest.PRIORITY_BALANCED_POWER_ACCURACY)
                                .setInterval(AppUtils.UPDATE_INTERVAL_LOCATION)
                                .setFastestInterval(AppUtils.FASTEST_INTERVAL_LOCATION);
        } else {
                if (isGPSEnabled)
                        locationRequest = LocationRequest.create()
                                .setPriority(LocationRequest.PRIORITY_HIGH_ACCURACY)
                                .setInterval(AppUtils.UPDATE_INTERVAL_GPS);
        }
                return locationRequest;
}
```

Just to remember, this *LocationRequest* is used as a parameter when requesting location updates.

Now, we have successfully set up our Location Updates. Every 5 or 20 seconds (depending if we need or not GPS) the *onStartCommand()* method of our Location Service will be called. This call is accompanied by an intent that contains the current location as an object of the class Location

```java
@Override
public int onStartCommand(Intent intent, int flags, int startId) {
        location=intent.getParcelableExtra(FusedLocationProviderApi.KEY_LOCATION_CHANGED);
        locationChanged(location);
        return START_NOT_STICKY;
}
```

### III.1.2. Activity Processing Algorithm

In order to receive and process activity recognition updates three methods are implemented. The first one is a broadcast receiver that obtains activity updates sent by the Activity Recognition Services every 5 seconds.

```java
private final BroadcastReceiver activityReceiver = new BroadcastReceiver() {
        @Override
        public void onReceive(Context context, Intent intent) {
                activityProcessing(intent.getStringExtra("activity"));
        }
};
```

The second one is the method used to count all the activity updates. Every time the sample is received we make a call to *activityProcessing()* method in order to count all the consecutive activity updates during the segment interval time (2 minutes).

The last one is used to return the most probable activity and it's called just after the 2 minutes have passed.

```java
private void activityProcessing(String _activity) {

        switch (_activity) {

                case "vehicle":
                        vehicle_count++;
                        break;
                case "bicycle":
                        bicycle_count++;
                        break;
                case "on_foot":
                        on_foot_count++;
                        break;
                case "still":
                        still_count++;
                        break;
                case "unknown":
                        unknown_count++;
                        break;

        }
}
```

```java
private String activityEstimation(int _isOnfoot, int _isBicycle, int _isStill, int _isVehicle, int _isUnknown) {

        ArrayList<Integer> max = new ArrayList<Integer>(5);
        // arraylist with size equal to the number of max
        max = findMax(_isOnfoot, _isBicycle, _isStill, _isVehicle, _isUnknown);
        switch (max.size()) {
        case 1:
                return (toActivity(max.get(0)));

        case 2:
           if (max.get(0) == 3 || max.get(1) == 3) { //if vehicle is inside the array->return vehicle
                        return toActivity(3);
                } else if (max.get(0) == 1 || max.get(1) == 1) { //same but for bicycle
                        return toActivity(1);
                } else if (max.get(0) == 0 || max.get(1) == 0) { //same but for on_foot
                        return toActivity(0);
                } else {
                        return toActivity(2); // else return -> still
                }
        case 3:
                if (max.get(0) == 3 || max.get(1) == 3 || max.get(2) == 3) {
                        return toActivity(3);
                } else if (max.get(0) == 1 || max.get(1) == 1 || max.get(2) == 1) {
                        return toActivity(1);
                } else {
                        return toActivity(0);
                }
        case 4:
                if (max.get(0) == 3 || max.get(1) == 3 || max.get(2) == 3 || max.get(3) == 3) {
                        return toActivity(3);
                } else {
                        return toActivity(1);
                }
        case 5:
                return toActivity(2); // if all samples equal 0 -> return still
                }
        return "still";
}
```

This method returns the activity with more number of samples. If we have two or more activities with the same number of maximum samples we have established a simple algorithm that only returns one.

This algorithm provides a priority order where vehicle has the maximum priority. If there are no vehicle samples, the priority order goes: bicycle, on_foot and finally still. Let's see some examples of this algorithm:

| Vehicle: 1 |
| On_foot: 5 |
| Bicycle: 0 |
| Still: 2 |
| Unknown: 1 |

The algorithm will return on_foot because is the activity with more samples.

| Vehicle: 5 |
| On_foot: 5 |
| Bicycle: 0 |
| Still: 5 |
| Unknown: 1 |

Now we have three activities with the same number of maximum samples. Vehicle has the maximum priority so the algorithm will return vehicle.

Also, we have implemented *toActivity()* method (used in previous code) to avoid making writing mistakes when returning the activity.

```java
private String toActivity(int index) {

        switch (index) {
        case 0:
                return "on_foot";
        case 1:
                return "bicycle";
        case 2:
                return "still";
        case 3:
                return "vehicle";
        case 4:
                return "unknown";
        }
        return "";
}
```

### III.1.3. Location Processing Algorithm

As we have already mentioned, we receive location updates every 20 seconds approximately. Every time this update arrives at *onStartCommand()* method, another one is called: *locationChanged().*
This is the most important method in MobilitApp so we will explain it in detail.
As it is large method it would be better to divide it in four stages.

#### III.1.3.1. First Stage: GPS activator algorithm

Sometimes GPS is needed because we are in a non-Wi-Fi zone or with bad Wi-Fi coverage. GPS uses so much battery power so it may impact in battery life when used in long-running applications like MobilitApp. For this reason, GPS will only be activated during two segment cycles (a maximum time of four minutes).

MobilitApp uses Google Fused Location Provider (as explained before), and its normal behaviour is to only use cell towers and Wi-Fi access points. When these two providers fail, GPS is needed but how can we notice it? There are two possibilities:

First, when there is no near Wi-Fi (e.g. highway, isolated towns) but we can still use cell towers locations. In this situation we receive location updates with little accuracy (i.e. ~700-1000 meters). So, when location accuracy is higher than 1000 meters GPS will be activated. But if the accuracy is much higher than 1000 meters (that means we are inside an underground), so there is no point enabling the GPS.

And second when we cannot get any location during the 20 seconds interval. When this happens, it sends repeatedly the last known location until a good one is received. So, if the same location is repeating GPS will be enabled.

The following code shows this algorithm:

```
if (location.getAccuracy() < 200) {
        tempLocationArrayList.add(location);
        // Checks whether the loc is the same as the previous --> Enable GPS
        if (!gpsON)
                if (tempLocationArrayList.size() != 1)
                  if (tempLocationArrayList.get(tempLocationArrayList.size() - 1).getLatitude()
                   == tempLocationArrayList.get(tempLocationArrayList.size() - 2).getLatitude()
                  && tempLocationArrayList.get(tempLocationArrayList.size() - 1).getLongitude()
                  == tempLocationArrayList.get(tempLocationArrayList.size() - 2).getLongitude())
                  {
                        reploc++;
                                if (reploc > 2) {
                                        if (vehicle_count > 1 || on_foot_count > 2) {
                                                needGPS(true);
                                                cada2min = 0;
                                                Log.v("aki", "aki_GPSON");
                                                gpsON = true;
                                        }
                                }
                } else {
                        reploc = 0;
                }
} else { // loc > 200
    if (location.getAccuracy() < 1000 && wifi.isWifiEnabled()) //GPS enabled when 200<loc<1000
        if (!gpsON) {
                needGPS(true);
                cada2min = 0;
                Log.v("aki", "aki_GPSON");
                gpsON = true;
        }

}
```

First part of the code enables GPS when we receive repeated locations and only when the detected activity is on_foot or vehicle, thereby avoid wasting battery power when still. Second part enables GPS when the accuracy is between 200 and 1000 meters.

To be able to enable or disable the GPS MobilitApp uses *needGPS()* method. When its parameter is true GPS will be enabled and vice versa.

```
private void needGPS(boolean value) {
        if (hasGPS) {
                Log.v("aki", "aki_needGPS");

                Requester req = new Requester(this);

                req.requestUpdates("LOC", value);
        }
}
```

As said before, Requester class is used in order to request new updates.

### III.1.3.2. Second Stage: Underground Recognition

This stage will be explained in Transport Recognition section.

### III.1.3.3. Third Stage: Mobility Data Parameters

In this stage we use the location samples stored in the ArrayList to obtain four important parameters: total and block distance, total duration and average speed.
All these parameters are computed right after the segment time (2 minutes).

Total distance is calculated from the latitudes and longitudes using the following formula:

```java
//Distance between two locations, in meters
public static double distanceCalculation(double firstLatitude, double firstLongitude, double lastLatitude, double lastLongitude) {

        double distance = 0;
        firstLatitude = Math.toRadians(firstLatitude);
        firstLongitude = Math.toRadians(firstLongitude);
        lastLatitude = Math.toRadians(lastLatitude);
        lastLongitude = Math.toRadians(lastLongitude);

        double x = (lastLongitude - firstLongitude) * Math.cos((firstLatitude + lastLatitude) / 2);
        double y = (firstLatitude - lastLatitude);

        distance = Math.sqrt(x * x + y * y) * AppUtils.EARTH_RADIUS;
        return distance;
}
```

The difference between total and block distance is that total distance is calculated between all the consecutive locations and the block one is only computed with the first and last location.
Block distance is used to consider still activity even the detected activity is on_foot when that distance is lower than the block size (e.g. movement inside our home, at work...).

```java
// Total Duration
segment_total_duration = (tempLocationArrayList.get(size - 1).getTime() -
tempLocationArrayList.get(0).getTime()) / AppUtils.MILLISECONDS_PER_SECOND;
// Block distance (not counting intermediate points) to compare with block size)
block_distance = AppUtils.distanceCalculation(tempLocationArrayList.get(0).getLatitude(),
tempLocationArrayList.get(0).getLongitude(), tempLocationArrayList.get(size - 1).getLatitude(),
tempLocationArrayList.get(size - 1).getLongitude());
// Total distance
for (int j = 0; j < size - 1; j++) {
        // Distance between consecutive locations, in meters
        segment_delta_distance = AppUtils.distanceCalculation(
        tempLocationArrayList.get(j).getLatitude(),
        tempLocationArrayList.get(j).getLongitude(),
        tempLocationArrayList.get(j + 1).getLatitude(),
        tempLocationArrayList.get(j + 1).getLongitude());
        // Sum of all distance_deltas of the segment
        total_segment_delta_distance += segment_delta_distance;
}
average_segment_speed = (total_segment_delta_distance * 3.6) / segment_total_duration;
```

### *III.1.3.4.    Fourth Stage: Creating Mobility Data Segments*

In this last stage we are going to create the mobility data segments. These segments include location points, activity type, distance, duration, speed and date of first and last location point.

Segment is a class created to store all mobility data in a way that can be easily processed, stored and parsed.

It is important to check whether the detected activity corresponds to what we are doing (e.g. detected activity is on_foot and the speed calculated is 40 km/h → wrong). For this reason a *switch-case* followed with *else-if* are implemented in this part of the code. *Switch* will select the current detected activity and with *else-if* the previous activity will be selected.

We have 5 possible activities and for each one 5 more depending on the previous one. It makes a total of 25 blocks of code similar to each other. Let's see only one example to not fill too many pages with similar code.

**Detected Activity:** Vehicle
**Previous Activity:** On Foot

We get the size of location *ArrayList* in order to know how many location samples we have, then we create an empty or not segment to start setting all its fields with mobility data. If previous and current activity is the same, there is no need to create a new segment but to update the previous segment adding the new location samples and updating the rest of parameters. To do that, temporally variables are created to store previous parameters (i.e. temp_speed, temp_duration and temp_distance).

This segment is added to a global *ArrayList* where all obtained segments are going to be stored.

```
case "vehicle":
// segment vehicle followed by on_foot, it's on_foot + vehicle
if (lastActivitiesArrayList.get(0).equals("on_foot")) {
        if (average_segment_speed > AppUtils.MAX_ON_FOOT_SPEED) {
                lastActivitiesArrayList.clear();
                lastActivitiesArrayList.add("vehicle");
                int segment_size = tempLocationArrayList.size();
                // create new segment to store the new values
                segment = new Segment();
                segment.setFirstLocation(tempLocationArrayList.get(0));
                segment.setLastLocation(tempLocationArrayList.get(segment_size - 1));
                segment.setTotalDistance(total_segment_delta_distance);

                segment.setTotalDuration(segment_total_duration);
                segment.setTotalAverageSpeed(average_segment_speed);
                segment.setActivity("vehicle");
                segment.updateLocationPoints(tempLocationArrayList);
                // add current segment
                segmentArrayList.add(segment);
                // new values to temp_variables
                temp_duration = segment_total_duration;
                temp_distance = total_segment_delta_distance;
                temp_speed = average_segment_speed;
                tmpLastLocation = tempLocationArrayList.get(segment_size - 1);
```

```java
} else {
                if (block_distance > AppUtils.BLOCK_RADIUS) {
                        lastActivitiesArrayList.clear();
                        lastActivitiesArrayList.add("on_foot");
                        int segment_size = tempLocationArrayList.size();

                        // update variables of current segment on_foot
                        temp_duration += segment_total_duration;
                        temp_distance += total_segment_delta_distance;
                        temp_speed = (temp_speed + average_segment_speed) / 2;

                        // update segment data
                        segment.setTotalDistance(temp_distance);
                        segment.setTotalDuration(temp_duration);
                        segment.setTotalAverageSpeed(temp_speed);
                        segment.setLastLocation(tempLocationArrayList.get(segment_size - 1));
                        segment.setActivity("on_foot");
                        segment.updateLocationPoints(tempLocationArrayList);

                        // update current segment
                        segmentArrayList.set(segmentArrayList.size() - 1, segment);
                        tmpLastLocation = tempLocationArrayList.get(segment_size - 1);
                } else {
                        lastActivitiesArrayList.clear();
                        lastActivitiesArrayList.add("still");
                        int segment_size = tempLocationArrayList.size();
                        segment = new Segment();

                if (segment_size > 1) {
                        tempPlace = tempLocationArrayList.get(0);
                for (int j = 0; j < segment_size; j++) {
                        if (tempPlace.getAccuracy() > tempLocationArrayList.get(j).getAccuracy())
                                tempPlace = tempLocationArrayList.get(j);
                        }
                }
                        segment.setFirstLocation(tempLocationArrayList.get(0));
                        segment.setLastLocation(tempLocationArrayList.get(segment_size - 1));
                        segment.setTotalDistance(0);
                        segment.setTotalDuration(segment_total_duration);
                        segment.setTotalAverageSpeed(0);
                        segment.setActivity("still");
                        segment.addSingleLocationPoint(tempPlace);

                        // new values to temp_variables
                        temp_duration = segment_total_duration;
                        temp_distance = 0;
                        temp_speed = 0;
                        segmentArrayList.add(segment);
                        tmpLastLocation = tempLocationArrayList.get(segment_size - 1);
                }
        }
}
```

As can be appreciated, average speed and block distance are checked before we can be sure it is on_foot, vehicle or still. In still activity we are going to add only the most accurate location sample and distance and speed will be set to zero.

## III.2.  Activity Recognition Service

Activity Recognition Service objective is to detect user's activity and send the result every 5 seconds back to the Location Service for further processing.

### III.2.1.  Starting Activity Recognition Service

This service is initialized right after the Location Service in the same way as this service, but instead of writing "LOC" we write "AR".

```
mrequesterAR.requestUpdates("AR", false);
```

With this line of code, a pending intent is launched every 5 seconds to our Activity Recognition class implemented as an *IntentService*. Google recommends using *IntentService* instead of Service but its performance will remain the same. The main difference is that in *IntentService* all the activity is handled in another thread and stops itself when finished.

### III.2.2.  Architecture

Every 5 seconds *onHandleIntent()* method is called, here we can extract and list all possible detected activities and finally get the most probable one.

```
@Override
protected void onHandleIntent(Intent intent) {

        // If the incoming intent contains an update
        if (ActivityRecognitionResult.hasResult(intent)) {

        // Get the update
        ActivityRecognitionResult result = ActivityRecognitionResult.extractResult(intent);

        // Get the most probable activity
        DetectedActivity mostProbableActivity = result.getMostProbableActivity();

        // Get an integer describing the type of activity

        activityType = mostProbableActivity.getType();
        activityName = getNameFromType(activityType);

        /*
         * At this point, you have retrieved all the information for the
         * current update. You can display this information to the user in a
         * notification, or send it to an Activity or Service in a broadcast
         * Intent.
         */
        Intent intentActivity = new Intent("activityRecognition");
        sendActivityLocationBroadcast(intentActivity, activityName);

        }
}
```

### III.2.3. Activity Detection

Since the activity type is in integer format, we use a switch-case method called *getNameFromType()* that returns a string with the detected activity name. These are the activity types can be detected:

- *on_foot:* Activity type returned if the citizen is either walking or running.

- *bicycle:* Activity type returned if the citizen is on a bicycle.

- vehicle: Activity type returned if the citizen is on a motor vehicle (e.g. car, motorbike, bus,…).

- still: Activity type returned if the citizen is not moving.

- unknown: Activity type returned if Activity Recognition API is not capable to estimate the actual activity.

```java
private String getNameFromType(int activityType) {
        switch (activityType) {
        case DetectedActivity.IN_VEHICLE:
                return "vehicle";
        case DetectedActivity.ON_BICYCLE:
                return "bicycle";
        case DetectedActivity.ON_FOOT:
                return "on_foot";
        case DetectedActivity.STILL:
                return "still";
        case DetectedActivity.UNKNOWN:
                return "unknown";
        case DetectedActivity.TILTING:
                // tilting changed to still
                return "still";
                }
        return "null";
}
```

Finally, we need to broadcast the activity detected to the Location Service using *sendBroadcast()* method.

```java
private void sendActivityLocationBroadcast(Intent _intent, String _activity) {

        _intent.putExtra("activity", _activity);
        sendBroadcast(_intent);
}
```

### III.3.  Map

Map needs to be added as a fragment. There are two ways to add a fragment within our *Activity* (i.e. window): dynamically or by code. In our case, the map is added by code as follows: We add a fragment element to the activity's layout file to define a *Fragment* object. The following code will attach a *MapFragment* [20] to the activity.

```
<fragment
      android:id="@+id/map"
      android:layout_width="match_parent"
      android:layout_height="match_parent"
      class="com.google.android.gms.maps.MapFragment" />
```

To work with the map, we will need to implement the *OnMapReadyCallback* interface and set an instance of the callback on the *MapFragment* using *getMapAsync()* method.

```
MapFragment mapFragment = (MapFragment) getActivity().getFragmentManager()
                                .findFragmentById(R.id.map);
mapFragment.getMapAsync(this);
```

The callback is triggered when the map is ready to be used and provides a non-null instance of *GoogleMap* [21]. We can use this *GoogleMap* object to set the view options for the map and add markers.

The following code will show the first part of the callback method *onMapReady().*

```
@Override
public void onMapReady(GoogleMap googleMap) {

      googleMap.setMyLocationEnabled(true);
      googleMap.getUiSettings().setCompassEnabled(true);
      googleMap.getUiSettings().setMapToolbarEnabled(false);
(...)
}
```

The Maps API offers different UI controls that can be set visible or invisible. With the above code we activate two controls and deactivate one.

First, we activate the My Location button; this button appears in the top right corner of the screen. When a user clicks the button, the camera animates to focus on the user's current location. Second we activate the compass; this compass appears in the top left corner.
Finally, we deactivate the map toolbar; this toolbar gives the user quick access to the Google Maps application and MobilitApp does not need this behaviour.

In the second part of the code we will set up MobilitApp behaviour when clicking the map or clicking an info window. Clicking an info window will bring up a window showing the mobility data (i.e. speed, distance, duration, activity, co2 saved and calories burned).
Info window will appear when the user clicks one of the different markers and shows the activity detected with the start and end time. When the map is clicked this window will disappear.

In the following page, we will see an example of a marker, an info window and the pop up window and then, the code that makes it possible.

*Design and implementation of an Android application to analyze mobility patterns of citizens*

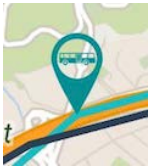

Figure 3.1: Marker

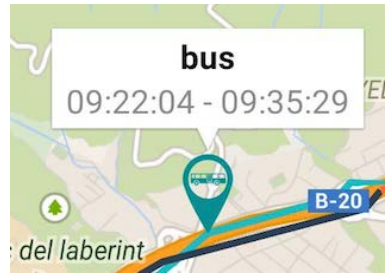

Figure 3.2: Info Window

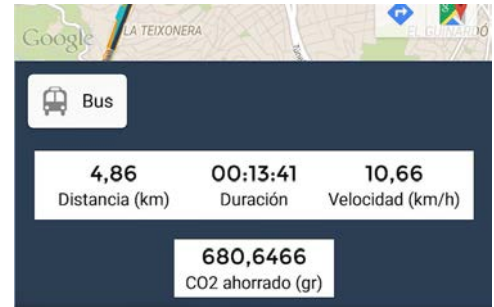

Figure 3.3: Pop-up Window

Before continuing with the above code, it's important to know how the markers and lines are drawn. When the user wants to display their mobility data on the map, MobilitApp loops through the file with the segment array with the method *readLocation()* (explained in History section) and in each loop *drawLocation()* is called.

```java
public void drawLocation(ArrayList<LatLng> _latlng, String _first_time, String _last_time, String
                _activity,String _line, double _distance, long _duration, double _speed, int index)
                throws IOException {
        Marker marker;
        int size = _latlng.size();
        if (!_latlng.isEmpty()) {
                if (first_location_history) {
                        first_location_history = false;
                        marker = googleMap.addMarker(new MarkerOptions()
                                .position(_latlng.get(0))
                                .title(_activity).snippet(_first_time + " - " + _last_time)
                                .icon(BitmapDescriptorFactory.fromResource(R.drawable.ic_still)));
                        segment = new Segment();
                        segment.setActivity(_activity);
                        segment.setTotalDistance(_distance);
                        if (_line != null) {
                                segment.setLine(_line);
                        }
                        segment.setTotalDuration(_duration);
                        segment.setTotalAverageSpeed(_speed);
                        segmentArray.add(index, segment);
                        hmap.put(marker, segment);
                } else {
                        if (_activity.equalsIgnoreCase("metro")) {
                                googleMap.addPolyline(new PolylineOptions().addAll(_latlng)
                                        .width(10).color(getMetroColor(_line)));
                        } else {
                                googleMap.addPolyline(new PolylineOptions().addAll(_latlng)
                                        .width(10).color(getColor(_activity)));
                        }
                        if (_activity.equals("still")) {
                                marker = googleMap.addMarker(new MarkerOptions()
                                        .position(_latlng.get(size - 1)).title(_activity)
                                        .snippet(_first_time + " - " + _last_time)
                                        .icon(BitmapDescriptorFactory
                                        .fromResource(R.drawable.ic_still)));
                                segment = new Segment();
                                segment.setActivity(_activity);
                                segment.setTotalDistance(_distance);
                                segment.setTotalDuration(_duration);
                                segment.setTotalAverageSpeed(_speed);
                                segmentArray.add(index, segment);
                                hmap.put(marker, segment);
```

```
            } else {
                    marker = googleMap.addMarker(new MarkerOptions()
                            .position(_latlng.get((size - 1) / 2)).title(_activity)
                            .snippet(_first_time + " - " + _last_time)
                            .icon(getActivityIcon(_activity)));
                    segment = new Segment();
                    segment.setActivity(_activity);
                    segment.setTotalDistance(_distance);
                    segment.setTotalDuration(_duration);
                    if (_line != null) {
                            segment.setLine(_line);
                    }
                    segment.setTotalAverageSpeed(_speed);
                    segmentArray.add(index, segment);
                    hmap.put(marker, segment);
}       }       }       }
```

Taking advantage of the fact we have the class Segment, we are going to use it to store the mobility parameters for later display when user clicks the info window.

To be able to know which information is shown when the user clicks in the different markers displayed on the map a *HashMap* [22] is created. A *HashMap* is a data structure consisting of a set of keys and values in which each key is mapped to a single value. In our case, the marker is the key and its value is the segment. So, when the user clicks in one of the markers on the map, the corresponding value (i.e. segment) will be selected.

Every segment has its own marker; these markers are situated just in the middle of the route and contain the snippet (i.e. the info window) and its corresponding icon depending on the activity detected.

Lines following the route are easily done with the method *addPolyline()*. We need locations in a specific format (i.e. using *LatLng* [23] class) and using *ArrayList*. Furthermore, we can set its width an colour.

Now that we know of the existence of the *HashMap*, we can continue with the code in the *onMapReady()* method. This second part of the code will show how to implement the action to view the mobility data when clicking the info window and how to make it invisible.

The view with all the mobility data is always there but it is invisible and only when the user clicks the info window becomes visible. When we want to make it invisible again, user only needs to click anywhere on the map.

First we need to get and convert the data into a more readable way: meters to kilometres and seconds to hours, minutes and seconds. Once we have all the data, we can fill the two *GridView* [24] using a custom adapter.

```
googleMap.setOnInfoWindowClickListener(new OnInfoWindowClickListener() {

        @Override
        public void onInfoWindowClick(Marker marker) {

        // Here we can get the mobility data, set the adapters and make the view VISIBLE

        }
};
```

As we can see, we have a reference of the marker in *onInfoWindowClick()* so, in order to get the mobility data we have to use the HashMap and use this marker as the key. For example:

```java
double speed = hmap.get(marker).getAverageSpeed();
```

We use the same custom adapter as in the Profile fragment; the only difference is that now the *TextView* [25] are in vertical rather than horizontal.
We have two *GridView* so we need two adapters, the first is used for the speed, distance and duration and the second is used only when the activity is on_foot (for calories burned) and bus, metro and train (for co2 saved).

```java
hlv.setAdapter(new AnotherAdapter(MainActivity.c, R.layout.dostv, valors, oho));
hlv2.setAdapter(new AnotherAdapter(MainActivity.c, R.layout.dostv, valors2, oho2));
```

Finally, we need to make visible the first *GridView* regardless of the activity detected, and the second one only when the activity is suitable with co2 saved and calories burned (i.e. on_foot, bus, metro and train). These two *GridView* are inside a *LinearLayout* [26] to be displayed correctly, so this one also needs to be set to visible.

```java
ll.setVisibility(View.VISIBLE);
hlv.setVisibility(View.VISIBLE);
```

To add the second *GridView* is necessary to move up the first one a little bit, so we are going to use margins.

```java
ViewGroup.MarginLayoutParams params = (MarginLayoutParams) hlv.getLayoutParams();
int dens = MainActivity.c.getResources().getDisplayMetrics().densityDpi;
if (hmap.get(marker).getActivity().equalsIgnoreCase("on_foot")
        || hmap.get(marker).getActivity().equalsIgnoreCase("metro")
        || hmap.get(marker).getActivity().equalsIgnoreCase("bus")
        || hmap.get(marker).getActivity().equalsIgnoreCase("renfe")) {

            params.setMargins(15 * dens / 160, 7 * dens / 160, 15 * dens / 160, 15 * dens / 160);
            hlv.setLayoutParams(params);
            hlv2.setVisibility(View.VISIBLE);
} else {
            params.setMargins(15 * dens / 160, 15 * dens / 160, 15 * dens / 160, 15 * dens / 160);
            hlv.setLayoutParams(params);
            hlv2.setVisibility(View.GONE);
}
```

Finally we need to set up the action of making invisible this information when user clicks on the map. For this purpose, we will use the *setOnMapClickListener()* method.

```java
googleMap.setOnMapClickListener(new OnMapClickListener() {

        @Override
        public void onMapClick(LatLng point) {

                ll.setVisibility(View.GONE);
                hlv.setVisibility(View.INVISIBLE);
                hlv2.setVisibility(View.INVISIBLE);
        }
});
```

## III.4. Login/Profile

MobilitApp lets user log in using Facebook or Gmail but it is not necessary since all the user's data is used locally and any information is used outside.

The reason for logging in is just to know the username to address him when MobilitApp needs user interaction.

Whenever user is logged in, profile window will appear where all user information is displayed.

User information is gathered when user clicks the login button and finally stored locally using *SharedPreferences* class. Profile window provides the user the possibility of logging out by clicking the Log Out button.

Login and Profile are two different fragments and only one is shown depending if the user is logged or not.

```
if (nombre.equalsIgnoreCase("noname")) {
        getSupportFragmentManager().beginTransaction()
                                .add(R.id.container_loginandperf, new LoginFragment(),
                                        "login").commit();
} else {
        getSupportFragmentManager().beginTransaction()
                                .add(R.id.container_loginandperf, new PerfilFragment(),
                                        "perfil").commit();
}
```

*Nombre* is the string where the username is stored, if the string is equal to "noname" then login fragment is added if not, profile fragment is the one added.

### III.4.1. Log in with Facebook

When user clicks Facebook button [27] a callback is generated and received in the login fragment:

```
private Session.StatusCallback callback = new Session.StatusCallback() {

        @Override
        public void call(Session session, SessionState state, Exception exception) {

                onSessionStateChange(session, state, exception);
        }
};
```

Now we need to gather user information to be later used in profile window, to be able to obtain this user information, we need read permissions.

```
authButton.setReadPermissions(Arrays.asList("public_profile", "email", "user_birthday"));
```

With these permissions we are able to obtain the following information: name, surname, gender, birthday and the e-mail.

Once the information is gathered and stored locally the session is closed and login fragment is finished.

```java
private void onSessionStateChange(Session session, SessionState state, Exception exception) {

if (state.isOpened()) {
        Request.newMeRequest(session, new Request.GraphUserCallback() {

                @Override
                public void onCompleted(GraphUser user, Response response) {
                        if (response != null) {
                                SharedPreferences.Editor edi = prefsFace.edit();
                                if (user.getBirthday() != null)
                                        edi.putString("birthday", user.getBirthday());
                                edi.putString("email",  user.asMap().get("email").toString());
                                edi.putString("nombre", user.getFirstName());
                                edi.putString("apellidos", user.getLastName());
                                edi.putString("genero", user.asMap().get("gender").toString());
                                edi.commit();
                        }

                }
        }).executeAsync();
                try {
                        Thread.sleep(200);
                } catch (InterruptedException e) {
                                e.printStackTrace();
                }
                Intent i = new Intent();
                getActivity().setResult(Activity.RESULT_OK, i);
                getActivity().finish();

        } else if (state.isClosed()) {
                session.closeAndClearTokenInformation();
}
}
```

### III.4.2. Log in with Google (Gmail account)

We need to create a Google client [28] to be able to log in with Google, the login is done with Google Plus using Plus API. Furthermore, we need permissions (also called scopes) to access different sections of user information.

```java
mPlusClient = new GoogleApiClient.Builder(getActivity(), this, this).addApi(Plus.API)
        .addScope(Plus.SCOPE_PLUS_LOGIN).build();
```

Once the Plus client is built we need a connection to gather all the information, store it, close session and finally finish login fragment. Connection is done with *connect()* method and if the connection is successful *onConnected()* method is called.

With Scope Plus login permission we can obtain the same information as before: e-mail, birthday, name, surname and gender.

```java
@Override
public void onConnected(Bundle connectionHint) {

        mConnectionProgressDialog.dismiss();

        p = Plus.PeopleApi.getCurrentPerson(mPlusClient);
        acc = Plus.AccountApi.getAccountName(mPlusClient);
        editorG = prefsGoogle.edit();

        editorG.putString("correo", acc);
        editorG.putString("birthday", p.getBirthday());
        editorG.putString("nombre", p.getName().getGivenName());
        editorG.putString("apellidos", p.getName().getFamilyName());
        editorG.putInt("genero", p.getGender());
        editorG.commit();
        signOut();
        Intent i = new Intent();
        getActivity().setResult(Activity.RESULT_OK, i);
        getActivity().finish();
}
```

*signOut()* method clears the default account and disconnects the client:

```java
public void signOut() {
        if (mPlusClient.isConnected()) {
                Plus.AccountApi.clearDefaultAccount(mPlusClient);
                mPlusClient.disconnect();
        }
}
```

### III.4.3. Profile Window

All the information gathered from login plus the weight obtained in preferences windows is shown in this window. The information is displayed using a *GridView* generated in the layout xml file with only one column. *GridView* needs an adapter to display the information; again, we will use a custom one so we can use two text views instead of one. In the first text view we will have the information name (i.e. name, surname, gender, age, weight and e-mail). And in the second one its corresponding information stored locally.

```java
titulo[0] = getActivity().getResources().getString(R.string.name) + "            ";
titulo[1] = getActivity().getResources().getString(R.string.surname) + "            ";
titulo[2] = getActivity().getResources().getString(R.string.gender) + "            ";
titulo[3] = getActivity().getResources().getString(R.string.age) + "            ";
titulo[4] = getActivity().getResources().getString(R.string.weight) + "            ";
titulo[5] = getActivity().getResources().getString(R.string.mail) + "            ";
```

Thanks to the Resources we can set the information name in other languages (i.e. English, Spanish and Catalan).

Creating a custom adapter is easy: First we need to extend the *ArrayAdapter* [29] class, next we create the constructor to pass the information and finally it is time to override *getView()* method.

In this method we set these two text views with the information passed in the constructor and we can change the typeface if we want.

```java
@Override
public View getView(int position, View convertView, ViewGroup parent) {
    View row = convertView;
    Typeface mFontreg = Typeface.createFromAsset(this.getContext().getAssets(),
            "fonts/Montserrat-Regular.ttf");
    Typeface mFontbold = Typeface.createFromAsset(this.getContext().getAssets(),
            "fonts/Montserrat-Bold.ttf");
    if (row == null) {
        LayoutInflater inflater = (LayoutInflater)
                context.getSystemService(Context.LAYOUT_INFLATER_SERVICE);
        row = inflater.inflate(layoutResourceId, parent, false);
    }
    TextView tv = (TextView) row.findViewById(R.id.text1);
    tv.setTypeface(mFontbold);
    TextView tv2 = (TextView) row.findViewById(R.id.text2);
    tv.setTypeface(mFontreg);
    tv.setText(files[position]);
    tv2.setText(files2[position]);
    return row;
}
```

Finally, we add this custom adapter to the *GridView*:

```java
gvprof.setAdapter(new AnotherAdapter(getActivity(), R.layout.dostvh, titulo, valores));
```

The layout passed as a parameter is the xml containing the two *TextView.*

## III.5. History

History is used to show users all generated files containing their mobility data. This mobility data is in list form and each row is named by its corresponding date (e.g. 10-02-2015).
Mobility data stored can be seen on the map as routes with different colours depending on the type of activity also, can be deleted locally.

### III.5.1. Structure

History window is created with a single fragment in which we can place our list of mobility data ordered by date. If the number of files containing that data is high, adding fragment normally could produce lag when clicking navigation drawer (left menu). To solve that, fragment will be added 100ms after using *AsyncTask* [30].
We need a handler to perform the populate action after 100ms. This handler is created is *onCreate()* method of History class.

```
h = new Handler();

h.postDelayed(new Runnable() {
        @Override
        public void run() {
                new Populate().execute();
        }
}, 100);
```

*Populate()* class adds history fragment in the current window.

```
public class Populate extends AsyncTask<Void, Void, Void> {

        @Override
        protected Void doInBackground(Void... params) {

                getSupportFragmentManager().beginTransaction().add(R.id.containerhist, new
                        HistoryFragment()).commit();
                return null;
        }
}
```

The view of the fragment is generated from an Android View called *RecyclerView* [31] and it is the flexible version of *ListView* [32]. This widget is a container for displaying large data sets that can be scrolled very efficiently by maintaining a limited number of views.
The main difference between other views is that as soon as a user scrolls a currently visible item out of view, this item's view can be recycled and reused whenever a new item comes into view.

*RecyclerView* needs to be set. The most important parameters are the adapter and the layout manager. The adapter provides access to the data items and the layout manager is responsible for measuring and positioning item views within the *RecyclerView.*

*RecyclerView* is set in *onActivityCreated()* method, this method is called when the fragment's activity has been created.

The following code shows how it is done:

```java
@Override
public void onActivityCreated(Bundle savedInstanceState) {
        super.onActivityCreated(savedInstanceState);

        readFolder();

        rV = (RecyclerView) getActivity().findViewById(R.id.recicler);
        rV.setHasFixedSize(true);
        rV.setAdapter(new HistoryAdapter(getActivity(), files, R.layout.customadap));
        rV.setLayoutManager(new LinearLayoutManager(getActivity()));
        rV.addItemDecoration(new DividerItemDecoration(getActivity(),
                DividerItemDecoration.VERTICAL_LIST));
}
```

*ReadFolder()* is the method used to read the folder where all mobility data files are stored. Instead of using a premade adapter like *ArrayAdapter*, MobilitApp uses a custom adapter named *HistoryAdapter*. Using a custom adapter adds flexibility to our adapter; therefore we can add a button used to delete files. This button will be visible when long-clicking the desired row.

When using *RecyclerView* is mandatory to use a *ViewHolder* [33] where all the row items will be initialized, in our case: the *TextView* where the date is shown and the button to delete.

```java
public static class ViewHolder extends RecyclerView.ViewHolder {
        public Button b;
        public TextView tv;

        public ViewHolder(View itemView) {
                super(itemView);

                tv = (TextView) itemView.findViewById(R.id.tvfiles);
                b = (Button) itemView.findViewById(R.id.bDelete);
        }
}
```

As commented previously, the key of *RecyclerView* class is that *View* can be recycled so initialization is done only once. This *ViewHolder* is needed in *onCreateViewHolder()* method because it is called when *RecyclerView* needs a new *ViewHolder* to represent an item.

```java
@Override
public ViewHolder onCreateViewHolder(ViewGroup parent, int i) {
        builder = new AlertDialog.Builder(context);
        View itemLayoutView = LayoutInflater.from(parent.getContext()).inflate(
                        idlay, parent, false);
        ViewHolder viewHolder = new ViewHolder(itemLayoutView);
        return viewHolder;
}
```

*OnBindViewHolder()* method is called by *RecyclerView* to display the data at the specified position. This method should update the contents of the *View* (above code) to reflect the item at the given position. Here we have the three listeners for the click and long-click events. When clicking we go back to the map and show the corresponding data and when long-clicking the delete button will be shown.

```java
@Override
public void onBindViewHolder(final ViewHolder vh, final int position) {

        vh.tv.setText(" "+" "+" "+data.get(position));
        vh.b.setVisibility(Button.INVISIBLE);
        vh.itemView.setOnClickListener(new OnClickListener() {
                @Override
                public void onClick(View v) {
                        Intent i = new Intent(context, MainActivity.class);
                        i.putExtra("archivo", data.get(position));
                        ((Activity) context).setResult(Activity.RESULT_OK, i);
                        ((Activity) context).finish();
                }
        });
        vh.itemView.setOnLongClickListener(new OnLongClickListener() {
                @Override
                public boolean onLongClick(View v) {
                        if (antbot != null)
                                antbot.setVisibility(Button.INVISIBLE);
                        antbot = vh.b;
                        vh.b.setVisibility(Button.VISIBLE);
                        return true;
                }
        });
        vh.b.setOnClickListener(new OnClickListener() {
                @Override
                public void onClick(View v) {
                        final File f = new File(ruta, data.get(position)
                                        + "location_segment.json");
                        if (System.currentTimeMillis() - f.lastModified() > 259200000) {
                                builder.setMessage(R.string.segur)
                                        .setTitle(R.string.borrar);
                                builder.setPositiveButton(R.string.acep,
                                        new DialogInterface.OnClickListener() {
                                        @Override
                                        public void onClick(DialogInterface dialog, int which) {
                                                f.delete();
                                                data.remove(position);
                                                notifyItemRemoved(position);
                                        }
                                });
                                builder.setNegativeButton(R.string.cancel, null);
                                builder.create();
                                builder.show();
                        } else {
                                builder.setMessage(R.string.rec).setTitle(R.string.borrar);
                                        builder.create().show();
                        }
                }
        });
}
```

MobilitApp sends the mobility data user wants to display to the map, which is in another window. To do this, *Intent* is created with the date of the file as an extra. Then, in the map window, *Intent* is read and can be properly shown to the user.

History window is launched when clicking the left menu in main window with *startActivityforResult()* and with request code 12. Therefore, when an intent is sent from History window *onActivityResult()* method catches it.

```java
@Override
protected void onActivityResult(int requestCode, int resultCode, Intent data) {

        switch (requestCode) {

        case 12:
                if (resultCode == Activity.RESULT_OK) {
                        frag.drawRoutes(data.getStringExtra("archivo"));
                }
                break;
        }
}
```

Finally, we need to call *drawRoutes()* method, which is situated in the fragment, with the extra data included in the intent.

```java
public void drawRoutes(String path) {

        googleMap.clear();
        readLocation(Environment.getExternalStorageDirectory().getAbsolutePath() +
                "/Download/MobilitApp/" + path+ "location_segment.json");
}
```

*ReadLocation()* method parses the JSON file and loops through the *JSONArray* [34] to draw one by one every route.

```java
JSONObject jsonObj = new JSONObject(jsonStr);
// Getting JSON Array node
segmentsData = jsonObj.getJSONArray("segments");
for (int i = 0; i < segmentsData.length(); i++) {
        JSONObject segment = segmentsData.getJSONObject(i);
        _first_time = segment.getString("first time");
        last_time = segment.getString("last time");
        _activity = segment.getString("activity");
        _distance = segment.getDouble("distance (m)");
        if (_activity.equalsIgnoreCase("metro")) {
                _line = segment.getString("line");
        }
        _duration = segment.getLong("duration (s)");
        _speed = segment.getDouble("speed (Km/h)");
        _latlngList = new ArrayList<LatLng>();
        if (_latlng != null) {
                _latlngList.add(_latlng);
        }
        JSONArray location_points = segment.getJSONArray("location");
        for (int j = 0; j < location_points.length();) {
                _latlng = new LatLng(location_points.optDouble(j), location_points.optDouble(j + 1));
                _latlngList.add(_latlng);
                j = j + 3;
        }
    drawLocation(_latlngList, _first_time, _last_time, _activity, _line, _distance, _duration, _speed, i);
}
```

Location parameters are used to draw the routes and the rest of parameters are used to display the information about the journey (i.e. speed, duration, distance...).

In order to draw the journeys, lines tracing the route are used. Android uses the class *Polyline* [35] to draw these lines between location points. In the Map section everything related to the Map will be explained.

```
if (_activity.equalsIgnoreCase("metro")) {
        googleMap.addPolyline(new PolylineOptions().addAll(_latlng).width(10)
                .color(getMetroColor(_line)));
} else {
        googleMap.addPolyline(new PolylineOptions().addAll(_latlng).width(10)
                .color(getColor(_activity)));
}
```

## III.6. Graph

Nowadays citizens are becoming increasingly aware about the environment and their health condition. Therefore, calculating the calories burned by physical activity and the amount of CO2 emissions saved might be a good to idea.
Formulas used are very simple and will be improved in future releases of MobilitApp.
All the calculations are done in post processing section (right before uploading the file) and stored locally using *SharedPreferences* class.

### III.6.1. Calories Burned Calculation

MobilitApp uses the metabolic equivalent formula to estimate the number of calories burned.

$$1 \text{ MET} \equiv 1\frac{\text{kcal}}{\text{kg} * h}$$

Figure 3.1: Metabolic Equivalent Formula

| | |
|---|---|
| walking, 1.7 mph (2.7 km/h), level ground, strolling, very slow | 2.3 |
| walking, 2.5 mph (4 km/h) | 2.9 |

Table 3.1: Equivalent MET ratio for walking (two different speeds)

Now that we know the ratios for walking activity, estimate the number of calories burned is easy. First we need to loop through the *JSONArray* looking for all on_foot activities and get its distance and speed parameters. Finally apply the formula and repeat the process until the end of the *JSONArray*.

```java
private void calperday(String day) {

        int jsonsize;
        float totcal = 0;
        if (!segmentArrayList.isEmpty()) {
                jsonsize = segmentArrayList.size();
                for (int i = 0; i < jsonsize; i++)
                        if (segmentArrayList.get(i).getActivity().equalsIgnoreCase("on_foot"))
                                totcal += calories(segmentArrayList.get(i).getTotalDuration(),
                                        segmentArrayList.get(i).getAverageSpeed());
                calo.edit().putFloat(day, totcal).commit();
        }
}
```

The formula is used in *calories()* method:

```java
private double calories(long duration, double speed) {

        double hours = duration / 3600.0;
        double parameter;
        if (speed < 2.7) {
                parameter = 2.3;
        } else {
                parameter = 2.9;
        }
        return parameter * Integer.parseInt(peso.getString("peso", "0")) * hours;
}
```

### III.6.2. CO2 Saved Calculation

The average CO2 emissions from car are 140 gr/km. As before, we need to loop through the *JSONArray* looking for all transport activities and get its distance parameter. Finally apply the formula and repeat the process until the end of the *JSONArray*.

```java
private void co2perday(String day) {

        int jsonsize;
        double totco2 = 0;

        if (!segmentArrayList.isEmpty()) {
                jsonsize = segmentArrayList.size();

                for (int i = 0; i < jsonsize; i++)
                        if (segmentArrayList.get(i).getActivity().equalsIgnoreCase("metro")
                        || segmentArrayList.get(i).getActivity().equalsIgnoreCase("bus")
                        || segmentArrayList.get(i).getActivity().equalsIgnoreCase("tram")
                        || segmentArrayList.get(i).getActivity().equalsIgnoreCase("renfe"))

                                totco2 += co2calc(segmentArrayList.get(i).getTotalDistance());

                co2.edit().putFloat(day, (float) totco2).commit();
        }
}
```

And *co2cal()* method applies the simple formula:

```java
private double co2calc(double distance) {

        double dist = distance / 1000.0;
        return 140 * dist;
}
```

### 1.1. Creating the graph

The graph has been created using a library made by Jonas Gehring [36]. MobilitApp uses this library since it is very flexible and easy to understand and implement.
The graph style used is the linear one:

```java
GraphView graphView = new LineGraphView(this, getResources().getString(R.string.graphtext));
```

We have two datasets, one for the calories burned and the other for the CO2 saved. Each dataset will be shown in different series with different colour.
The information to fill the dataset is stored locally as a shared preference.

```java
for (int i = 0; i < size; i++)
        for (int j = 0; j < size; j++) {
                if (label_calo.toArray()[j].toString().contains(labels[i])) {
                        datacal[i] = new GraphViewData(i, calo.getFloat((String)
                                label_calo.toArray()[j], 0));
                        dataco2[i] = new GraphViewData(i, co2.getFloat((String)
                                label_calo.toArray()[j], 0));
                }
        }
```

Now, we need to add these two series to the *GraphView*:

```
graphView.addSeries(new GraphViewSeries("Calorías (kcal)", new
        GraphViewSeriesStyle(Color.GREEN, 3), datacal));
graphView.addSeries(new GraphViewSeries("Co2 (gr)", new GraphViewSeriesStyle(Color.RED, 3),
        dataco2));
```

We can also make some extra customizations like: adding the legend, select the view port (i.e. the visible part) and make graph scrollable and scalable:

```
graphView.setShowLegend(true);
graphView.setLegendAlign(LegendAlign.MIDDLE);
graphView.getGraphViewStyle().setLegendWidth(300);

graphView.setViewPort(0, 7);
graphView.setScrollable(true);
graphView.setScalable(true);
```

Finally, it is time to add the *GraphView* generated to the current window:

```
LinearLayout layout = (LinearLayout) findViewById(R.id.gl);
layout.addView(graphView);
```

## III.7. Segment Files

In this chapter we are going to explain how it is structured the Segment class, how to write segments into a JSON file and finally an example of a JSON file.

### III.7.1. Segment Class

Segment class has the following variables:

```java
private Location _first_location;
private Location _last_location;
private double _total_distance;
private long _total_duration;
private double _total_average_speed;
private String _activity, _line;
private ArrayList<Location> _location_points = new ArrayList<Location>();
```

Then we have all the get and put method to obtain or modify the above variables. For example:

```java
//getting Last Location
public Location getLastLocation() {
        return this._last_location;
}
//setting Last Location
public void setLastLocation(Location last_location) {
        this._last_location = last_location;
}
//getting Location points
public ArrayList<Location> getLocationPoints() {
        return this._location_points;
}
//updating Location points
public void updateLocationPoints(ArrayList<Location> location_points) {
        this._location_points.addAll(location_points);
}
//update single Location point
public void addSingleLocationPoint(Location location) {
        this._location_points.add(location);
}
public Location getSingleLocationPoint() {
        return this._location_points.get(0);
}
```

### III.7.2. Writing JSON files

Mobility data is stored in an *ArrayList* of Segments and every time the segment is filled (i.e. every 2 minutes) this *ArrayList* is stored in external storage. File is not in "append mode" because we write the whole *ArrayList* instead of each segment (one below the other). This is because we want to store a single activity when we have two identical activities. So, instead of having "vehicle-vehicle" we will have only one "vehicle" with updated parameters.

*WriteSegment()* is the method that allow us to write JSON files.

```java
public void writeSegment(String _filename, ArrayList _segmentList) {
    try {
        JSONArray segments = new JSONArray();
        for (int k = 0; k < _segmentList.size(); k++) {
            // create the objects that will be in each spot of the segment array
            JSONObject single_segment = new JSONObject();
            single_segment.put("activity", _segmentList.get(k).getActivity());
            single_segment.put("distance (m)", _segmentList.get(k)
                    .getTotalDistance());
            single_segment.put("duration (s)", _segmentList.get(k)
                    .getTotalDuration());
            if (_segmentList.get(k).getActivity().equalsIgnoreCase("metro"))
                single_segment.put("line", _segmentList.get(k).getLine());

            single_segment.put("speed (Km/h)", _segmentList.get(k)
                    .getAverageSpeed());
            single_segment.put("first time", AppUtils.getTime(_segmentList.get(k)
                    .getFirstLocation().getTime()));
            single_segment.put("last time", AppUtils.getTime(_segmentList.get(k)
                    .getLastLocation().getTime()));
            // create an array inside the object for each spot of the segment array
            JSONArray location_points = new JSONArray();

            for (int j = 0; j < _segmentList.get(k).getLocationPoints().size(); j++) {
                location_points.put(_segmentList.get(k)
                        .getLocationPoints().get(j).getLatitude());
                location_points.put(_segmentList.get(k)
                        .getLocationPoints().get(j).getLongitude());
                location_points.put(AppUtils.getTime(_segmentList.get(k)
                        .getLocationPoints().get(j).getTime()));
            }
            single_segment.put("location", location_points);
            segments.put(k, single_segment);
        }
        JSONObject json = new JSONObject();
        json.put("segments", segments);
        try {
            File directory = new File(ruta);
            if (!directory.exists()) {
                directory.mkdir();
            }
            FileWriter fw = new FileWriter(_filename, false);
            fw.append(json.toString(1));
            fw.flush();
            fw.close();
        } catch (IOException e) {
            e.printStackTrace();
        }

    } catch (JSONException e) {
        e.printStackTrace();
    }
}
```

In JSON, we can use two different classes, *JSONObject* [37] and *JSONArray*. A *JSONObject* a modifiable set of name/value pairs while a *JSONArray* is an indexed sequence of values. In both cases, these values may be any mix of *JSONObject*, *JSONArray*, Strings, Booleans, Integers, Longs, Doubles and null.

Our JSON files will be written following the structure of the Segment class we have designed.

First, we have to loop through the ArrayList of Segments and using all the put methods (seen before), put field by field all information of each segment into a *JSONObject*.

The location field contains an array of multiple location samples with format of latitude, longitude and time. For this reason, location will be stored using a *JSONArray*.

Finally, we will put all *JSONObject*, each one of them with mobility data from different segments, into another *JSONArray*. This array will contain all the obtained segments and, at the same time, it will facilitate its further parsing and processing.

Once we have all the data stored in the array, we proceed to write it to a file. To write a file in Android we just need to use the *FileWriter* [38] class and use its methods: append, flush and close.

This *writeSegment()* method is going to be used in Location Service.

### III.7.3. JSON File Example

```
{
"segments": [
  {
   "activity": "on_foot",
   "distance (m)": 49.69776445992602,
   "duration (s)": 142,
   "speed (Km\/h)": 1.2599432,
   "first time": "09:46:44",
   "last time": "09:49:07",
   "location": [
    41.441145,
    2.1659081,
    "09:46:44",
    41.4410568,
    2.1660705,
    "09:47:11",
    41.441012,
    2.1661082,
    "09:47:32",
    41.4409738,
    2.1661926,
    "09:48:13",
    41.440959,
    2.1662142,
    "09:48:34",
    41.4410113,
    2.1663986,
    "09:49:07"
   ]
  },

  (...)

 ]
}
```

# Part IV
# Recognition Algorithms

## IV.1. Transport Recognition

In this section we are going to explain how MobilitApp recognizes the type of public transport citizen is using. MobilitApp can distinguish between: metro, bus, tram and train. For this purpose two Google APIs will be used: Directions API and Places API. These two APIs are explained in its respective section but to summarize, Places API is a service that returns information about places (e.g. establishments, metro stations...) using HTTP requests and Directions API can search for directions for several modes of transportation.

Recognition part is implemented inside Location Service. Metro recognition is done at the same time we receive location updates, the rest of modes of transportations are done in post processing section. This section is done right before the file is uploaded to the server.

Apart from these two Google APIs, MobilitApp uses a database of train and metro stations to complement the algorithms. These two databases are in JSON format and contain the name of the stations, their location, etc. Let's see an example of these two databases that will be explained later.

```
"metro":[
    {
     "id":"433",
     "line":"L5",
     "name":"Cornellà Centre",
     "connections":"T1-T2-T3-R.1-R4",
     "lat":"41.357315951168",
     "lon":"2.07044325855937"
    },
    {
     "id":"432",
     "line":"L5",
     "name":"Gavarra",
     "connections":"",
     "lat":"41.3580221811863",
     "lon":"2.07929676467967"
    },
 ...
```

```
"renfe":[
    {
     "line":"R16",
     "lat":"40.5958608323863",
     "lon":"0.449509090610943"
    },
    {
     "line":"R16",
     "lat":"40.8071840875766",
     "lon":"0.522786433876677"
    },
    {
     "line":"R16",
     "lat":"40.7590269187727",
     "lon":"0.556463577981504"
    },
 ...
```

### IV.1.1. Metro Recognition

Metro Recognition algorithm uses Places API to find the nearest station. Before explaining the algorithm, it is important to know that inside the metro, location accuracy will be higher than 1000 meters because there is neither Wi-Fi spots nor sky visibility.
This simple algorithm works as follows: when citizen enters a metro station Places API will find the nearest station. Once the metro journey is done and citizen comes to the surface, Places will find again the nearest station. Now we have origin and destination stations, we can trace the route following the metro line. This algorithm also allows up to two metro transfers.

```java
if (wifi.isWifiEnabled() && location.getAccuracy() > 1000) // loc accuracy > 1000 --> Underground
        badloc++;
```

Whenever location accuracy is higher than 1000 meters and Wi-Fi is activated a global variable named *badloc* is increased by one.

```java
if (badloc > 4 && goodloc != null && !afterbad) {
        write = false;
        loc1 = goodloc;
        afterbad = true;
        block = false;
}
```

Once the segment block time has passed, *badloc* variable is checked and if it is greater than 4 the following variables are set:

First, we set *write* to false so all the activity processing is not done.

Second, the last good location (before the bad ones) is stored.

Third, we set *block* to false. Block variable is used to avoid getting bad locations as good ones.

Finally *afterbad* variable is set to true to activate the following code:

```java
if (afterbad) {
        locgood.add(goodloc);
        if (locgood.size() > 1) // checks whether after bad locs good ones are received but
                                // repeated, if this happens we continue as before
                if (locgood.get(locgood.size() - 1).getLatitude() == locgood.get(locgood.size() -
                        2).getLatitude()&& locgood.get(locgood.size() - 1).getLongitude() ==
                                locgood.get(locgood.size() - 2).getLongitude()) {
                        locgood.remove(locgood.size() - 1);
                }
        if (locgood.size() > 2) {
                avoid = true;
                loc2 = locgood.get(0);
                badloc = 0;
                afterbad = false;
                tempLocationArrayList.clear();
                for (int index = 0; index < locgood.size(); index++)
                        tempLocationArrayList.add(locgood.get(index)); // clear previous
                                                            //locations and add new ones
                locgood.clear();
                try {
                        new GetPlaces("subway_station").execute(
                                new Double[] { loc1.getLatitude(), loc1.getLongitude(),
                                        loc2.getLatitude(), loc2.getLongitude() }).get();
                } catch (InterruptedException | ExecutionException e) {
                        // TODO Auto-generated catch block
                        e.printStackTrace();
                }
        }
}
```

It is important to always check if we are receiving repeated locations because it is common in bad location areas so, first part of the code checks that. In the second part of the code we are receiving good locations after bad ones so, it is supposed we have left the metro.

In this situation, we have origin and destination location and Places API will be used in *GetPlaces()* class.

Places API does HTTP requests to obtain places so it is mandatory to extend the class *AsyncTask* to our *GetPlaces* class. *AsyncTask* allows to perform background operations and publish results on the UI thread.

The following code will show how this class is implemented:

```java
private class GetPlaces extends AsyncTask<Double, Void, Void> {
        private String places;
        public GetPlaces(String places) {
                this.places = places;
        }
        @Override
        protected Void doInBackground(Double... loc) {
                PlacesService service = new PlacesService(API);
                ArrayList<String> result = service.findPlaces(loc[0], loc[1], loc[2], loc[3], places);
                if (result != null) {
                        if (result.size() == 2) {
                                orig = result.get(0);
                                dest = result.get(1);
                                if (!orig.equalsIgnoreCase(dest)) {
                                        getDistance(orig, dest);
                                } else {
                                        falsepositive = true;
                                }
                        } else {
                                falsepositive = true;
                        }
                } else {
                        falsepositive = true;
                }
                return null;
        }
}
```

*FindPlaces()* is the method that uses Places API (this method is explained in its respective section). It makes two http requests and depending on the replies may return three different results.
The first one is that the two requests return non-null result, it means Places API has found origin and destination station and returns the name of this two stations. The second one is that only one reply contains a string with the name of the station and the third one is that any reply contains a string. When we obtain the last two results *falsepositive* variable is set to true to leave this "metro" state.

Recognition is done when we obtain two different strings indicating origin and destination stations in that case, we can proceed finding the route with *getDistance()* method.
This method loops through the stations database and trace the route between origin and destination. The used algorithm can process up to one metro transfer. To differentiate different lines another parameter is added in the segment, this parameter is the metro line. Doing that, we will be able to change the line colour when tracing the route.

There is another database implemented as a String Array that only contains the name of the stations in the corresponding order the metro follows. This database of names will be used to search in the other database and find its corresponding location.

The algorithm used in *getDistance()* works as follows: we proceed from the fact we have origin and destination stations so the first thing we should do is to check if the two stations are on the same line. If so, the next step is to find the name of the origin station in the JSON database and get its corresponding location points. Then, loop through the second database to get the name of the following station and repeat the process until we reach the destination station.

The following example shows how it works in a short metro journey.

Metro L3 journey:
- **Origin**-> Canyelles
- **Destination**-> Montbau

```xml
<string-array name="L3">
    <item>Trinitat Nova</item>
    <item>Roquetes</item>
    <item>Canyelles</item>
    <item>Valldaura</item>
    <item>Mundet</item>
    <item>Montbau</item>
    <item>Vall d'Hebron</item>
    <item>Penitents</item>
    <item>Vallcarca</item>
    <item>Lesseps</item>
    <item>Fontana</item>
    <item>Diagonal</item>
    <item>Passeig de Gràcia</item>
    <item>Catalunya</item>
    <item>Liceu</item>
    <item>Drassanes</item>
    <item>Paral.lel</item>
    <item>Poble Sec</item>
    <item>Espanya</item>
    <item>Tarragona</item>
    <item>Sants Estació</item>
    <item>Plaça del Centre</item>
    <item>Les Corts</item>
    <item>Maria Cristina</item>
    <item>Palau Reial</item>
    <item>Zona Universitària</item>
</string-array>
```

```json
{
"id":"387",
"line":"L3",
"name":"Canyelles",
"connections":"",

"lat":"41.4417702469479",

"lon":"2.16633743592508"
},
...
{
"id":"386",
"line":"L3",
"name":"Valldaura",
"connections":"",

"lat":"41.4380515005838",

"lon":"2.15693039004997"
},
...
{
"id":"385",
"line":"L3",
"name":"Mundet",
"connections":"",

"lat":"41.4357688982928",

"lon":"2.14859090896243"
},
...
{
"id":"365",
"line":"L3",
"name":"Montbau",
"connections":"",

"lat":"41.4306544208465",

"lon":"2.14505193131497"
},
...
```

```json
{
"activity": "metro",
"distance (m)": 2270.152587890625,
"duration (s)": 410,
"line": "L3",
"speed (Km\/h)": 19.933047113185975,
"first time": "18:14:52",
"last time": "18:20:02",
"location": [
 41.4417702469479,
 2.16633743592508,
 "18:14:52",
 41.4380515005838,
 2.15693039004997,
 "18:14:52",
 41.4357688982928,
 2.14859090896243,
 "18:14:52",
 41.4306544208465,
 2.14505193131497,
 "18:14:52"
 ]
},
```

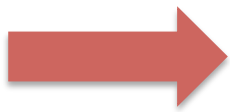

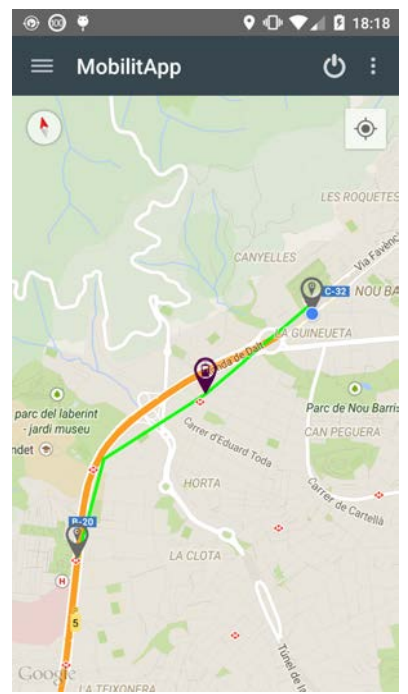

However, if origin and destination are not on the same line we should apply the algorithm twice as follows.

Metro journey between **Canyelles** (L3) and **Camp de l'Arpa** (L5), there are different ways/combinations to do this journey but the algorithm calculates all the combinations and return the shortest one, just as the citizen would.

So, first the algorithm is used between **Canyelles** and **Vall d'Hebrón** and finally between **Vall d'Hebrón** and **Camp de l'Arpa**.

Figure 4.1 shows this journey

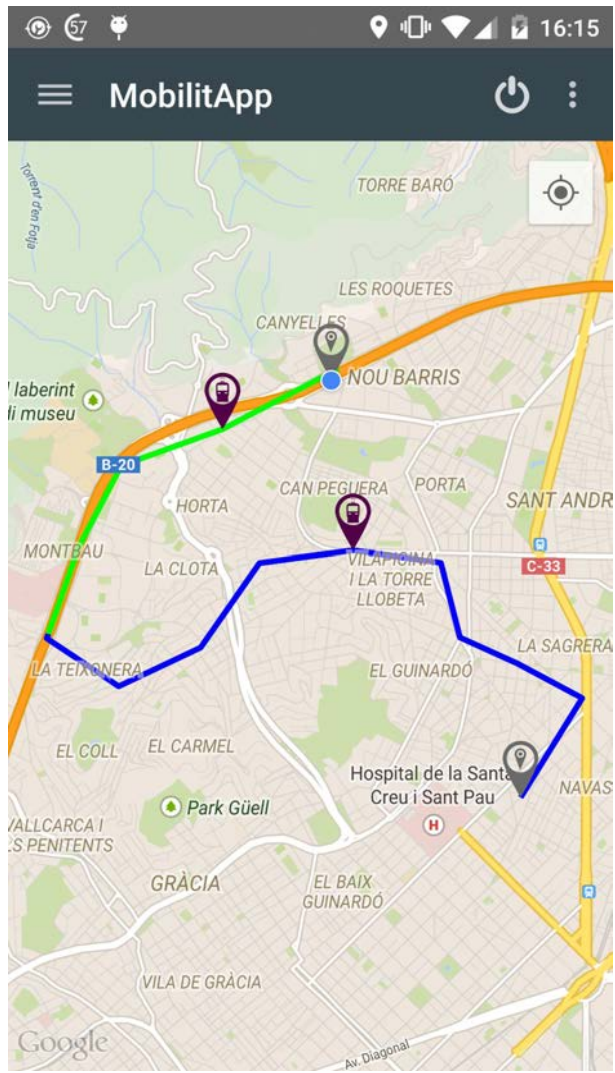

Figure 4.1: Metro Journey (Canyelles-Camp de l'Arpa)

### IV.1.2. Bus and Tram Recognition

Bus and Tram recognition use Google Directions API. Recognition is done right before uploading the file to the server.

Recognition algorithm seeks for vehicle activities and using Directions API tries to find bus or tram routes between beginning and end points taking into account departure time. So, maybe route between these two points exists but bus/tram departure time did not match start time of the vehicle sample.

```java
for (int i = 0; i < siz; i++) { // BUS/TRAM detection
        String t;
        if (segmentArrayList.get(i).getActivity().equalsIgnoreCase("vehicle")) {
                if (segmentArrayList.get(i - 1).getActivity().equalsIgnoreCase("still")) {
                        // if the previous sample is still we get the first vehicle sample
                        o = segmentArrayList.get(i).getFirstLocation();
                } else {
                        // else we get this last sample instead of the following as the beginning sample
                        o = segmentArrayList.get(i - 1).getFirstLocation();
                }
                d = segmentArrayList.get(i).getLastLocation();

                try {
                        t = new GetBusorTram((o.getTime()) / 1000, ((d.getTime()) /
                                1000).execute(new Double[] { o.getLatitude(), o.getLongitude(),
                                        d.getLatitude(),  d.getLongitude() }).get();

                        if (t.equalsIgnoreCase("BUS")) {
                                segmentArrayList.get(i).setActivity("bus");
                                        continue;
                        } else if (t.equalsIgnoreCase("TRAM")) {
                                segmentArrayList.get(i).setActivity("tram");
                                        continue;
                        }
                } catch (InterruptedException | ExecutionException e) {
                        e.printStackTrace();
                }
        }
}
```

Bus or Tram accelerates slowly so maybe the previous activity is not still (we are waiting). If so, we consider this last activity (normally is on_foot) as the first sample or origin of the bus journey. With this consideration we can now call the method *GetBusorTram()* that uses the Directions API. Directions API also makes http requests so it is mandatory to extend AsyncTask class.

```java
private class GetBusorTram extends AsyncTask<Double, Void, String> {
        long departure_time, arrival_time;

        public GetBusorTram(long departure_time) {
                this.departure_time = departure_time;
                this.arrival_time = arrival_time;
        }
        @Override
        protected String doInBackground(Double... params) {

                DirectionsService ds = new DirectionsService(params[0], params[1], params[2],
params[3], departure_time, arrival_time);
                return ds.findBusorTram();
        }
}
```

This class is explained in Directions API section, but to summarize it may return three strings: "none", "BUS" or "TRAM". If it is bus or tram we simply need to change the segment activity name with *setActivity()*.

### IV.1.3. Train Recognition

Train Recognition is done right after Bus/Tram recognition if "none" is returned. Places API does not return information about train stations so the algorithm does not use any Google API.

The algorithm finds the nearest station for both origin and destination; if that distance is below 50 meters and both stations belong the same line vehicle activity is changed to train activity. This is a simple algorithm that will be modified in future releases.

```java
private boolean nearstation(double lat, double longi) {

        double min = 100000;
        double dist;
        int min_index = 0;
        twopositionarray = new ArrayList<>(2);

        getRenfeLocs();

        for (int i = 0; i < arrayrenfe.size(); i++) {
                dist = AppUtils.distanceCalculation(arrayrenfe.get(i).latitude,
                        arrayrenfe.get(i).longitude, lat, longi);
                if (dist < min) {
                        min = dist;
                        min_index = i;
                }
        }
        if (min < 100) {
                twopositionarray.add(arrayrenfeline.get(min_index));
                return true;
        } else {
                return false;
        }
}
```

*getRenfeLocs ()* method loads the trains station database previously commented.

# Part V
# References

# References

[0]  Kantar Worldpanel. *Smartphone OS market share*.
http://www.kantarworldpanel.com/global/smartphone-os-market-share/

[1]  Adrian Latorre and Marc Llahona. *Project CICYT "RESPUESTA A EMERGENCIAS EN COMUNIDADES INTELIGENTES (EMRISCO)". TEC2013-47665-C4-1-R.*
http://sertel.upc.edu/~maguilar/simulators.html

[2]  Android Developers. *Activities*.
http://developer.android.com/guide/components/activities.html

[3]  Android Developers. *Fragments*.
http://developer.android.com/guide/components/fragments.html

[4]  Android Developers. *FragmentTransaction*.
http://developer.android.com/reference/android/app/FragmentTransaction.html

[5]  Android Developers. *Services*.
http://developer.android.com/guide/components/services.html

[6]  Android Developers. *Supporting Different Languages*.
http://developer.android.com/training/basics/supporting-devices/languages.html

[7]  Android Developers. *Storage Options*.
http://developer.android.com/guide/topics/data/data-storage.html

[8]  Android Developers. *SharedPreferences*.
http://developer.android.com/reference/android/content/SharedPreferences.html

[9]  Google. *Google Cloud Storage*. https://cloud.google.com/storage/?hl=en

[10] RailsGuides. *Session ID*. http://guides.rubyonrails.org/security.html#session-id

[11] Android Developers. *Location Request*.
https://developer.android.com/reference/com/google/android/gms/location/LocationRequest.html

[12] Android Developers. *ActivityRecognitionResult*.
http://developer.android.com/reference/com/google/android/gms/location/ActivityRecognitionResult.html

[13] Android Developers. *DetectedActivity*.
http://developer.android.com/reference/com/google/android/gms/location/DetectedActivity.html

[14] Google Developers. *Places API*.
https://developers.google.com/places/documentation/?hl=en

[15] Google Developers. *Directions API*.
https://developers.google.com/maps/documentation/directions/?hl=en

[16] Android Developers. *ArrayList*.
http://developer.android.com/reference/java/util/ArrayList.html

[17] Google Developers. *GoogleApiClient.Builder*.
http://developer.android.com/reference/com/google/android/gms/common/api/GoogleApiClient.Builder.html

[18] Android Developers. *GoogleApiClient*.
http://developer.android.com/reference/com/google/android/gms/common/api/GoogleApiClient.html

[19] Android Developers. *PendingIntent*.
http://developer.android.com/reference/android/app/PendingIntent.html

[20] Android Developers. *MapFragment*.

http://developer.android.com/reference/com/google/android/gms/maps/MapFragment.html

[21] Android Developers. *GoogleMap*.
http://developer.android.com/reference/com/google/android/gms/maps/GoogleMap.html

[22] Android Developers. *HashMap*.
http://developer.android.com/reference/java/util/HashMap.html

[23] Android Developers. *LatLng*.
http://developer.android.com/reference/com/google/android/gms/maps/model/LatLng.html

[24] Android Developers. *GridView*.
http://developer.android.com/reference/android/widget/GridView.html

[25] Android Developers. *TextView*.
http://developer.android.com/reference/android/widget/TextView.html

[26] Android Developers. *LinearLayout*.
http://developer.android.com/reference/android/widget/LinearLayout.html

[27] Facebook Developers. *Facebook Login for Android*.
https://developers.facebook.com/docs/android/login-with-facebook/v2.2

[28] Google Developers. *Google+ Sign-in for Android*.
https://developers.google.com/+/mobile/android/sign-in?hl=en

[29] Android Developers. *ArrayAdapter*.
http://developer.android.com/reference/android/widget/ArrayAdapter.html

[30] Android Developers. *AsyncTask*.
http://developer.android.com/reference/android/os/AsyncTask.html

[31] Android Developers. *RecyclerView*.
http://developer.android.com/reference/android/support/v7/widget/RecyclerView.html

[32] Android Developers. *ListView*.
http://developer.android.com/reference/android/widget/ListView.html

[33] Android Developers. *RecyclerView.ViewHolder*.
http://developer.android.com/reference/android/support/v7/widget/RecyclerView.ViewHolder.html

[34] Android Developers. *JSONArray*.
http://developer.android.com/reference/org/json/JSONArray.html

[35] Android Developers. *Polyline*.
http://developer.android.com/reference/com/google/android/gms/maps/model/Polyline.html

[36] Jonas Gehring. *GraphView*.  http://www.android-graphview.org/

[37] Android Developers. *JSONObject*.
http://developer.android.com/reference/org/json/JSONObject.html

[38] Android Developers. *FileWriter*.
http://developer.android.com/reference/java/io/FileWriter.html

# Part VI
# Annexes

# Annex A

# Journey Examples

## A.1.    Journey 1

**Information:** The journey begins at 6:30 and ends at 13:30 and two types of transport have been used: car and bus (two different lines).

We should remember that Android recognizes only on_foot, bicycle, still, vehicle and unknown activities so, we need to process the data to recognize more types of transport. In a first approach we should notice only vehicle activities even when using the bus or metro and then, after the post processing and using Google APIs, vehicle activities will change depending on the transport detected.

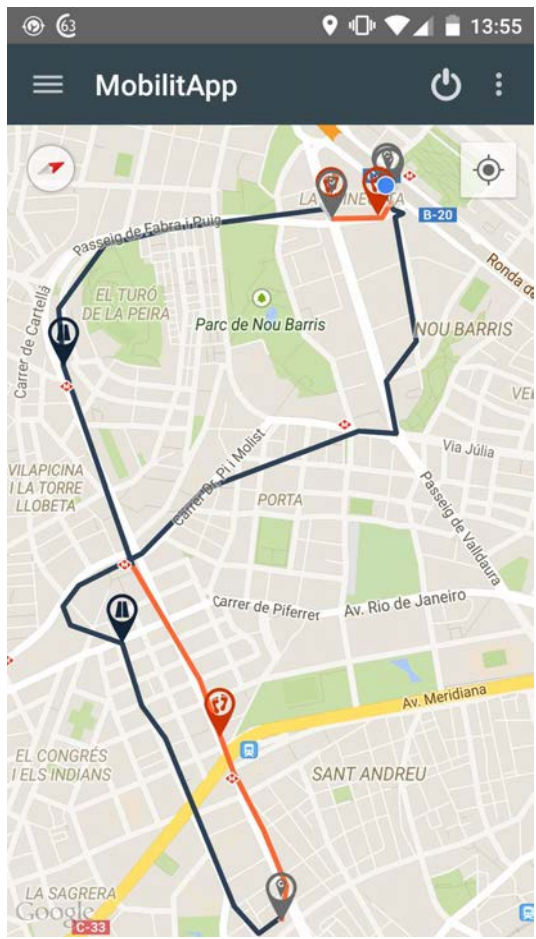
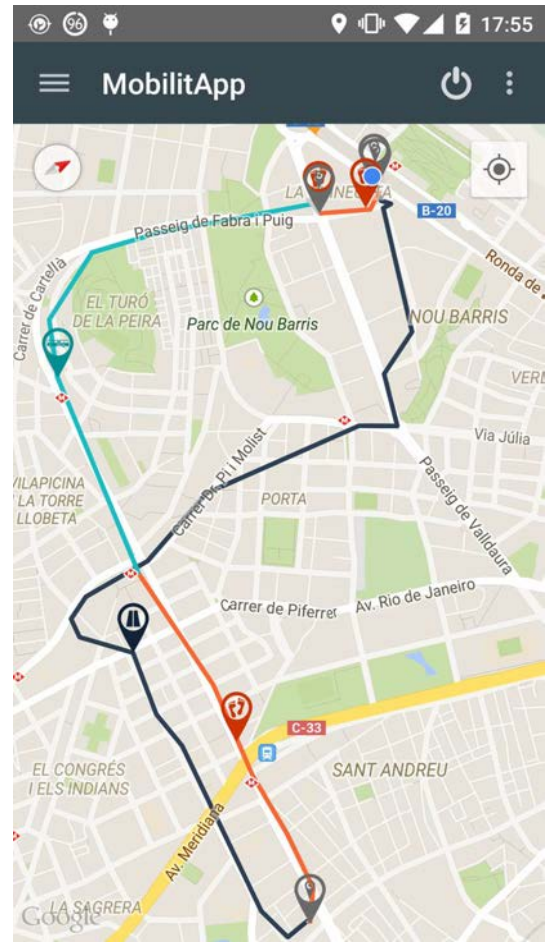

Left image shows the "raw" version, where we only notice vehicle activities, in the image on the right a vehicle activity has turned into bus activity.
The journey by car (black line in the image on the right) has been a total success, the route followed is almost the same and did not turn into bus activity because there was no route between origin and destination with these start and end time.

Conversely, only one of the two journeys by bus has been detected. As we can see, on_foot activity is detected instead of vehicle. This is due to a traffic jam in this zone so low speeds were detected causing this error.

## A.2.    Journey 2

**Information:** The journey begins at 8:05 and ends at 14:37 and the type of transport used is the Metro (round trip). The Metro ride is between Canyelles and Lesseps (Line 3) and after a long walk destination is reached.

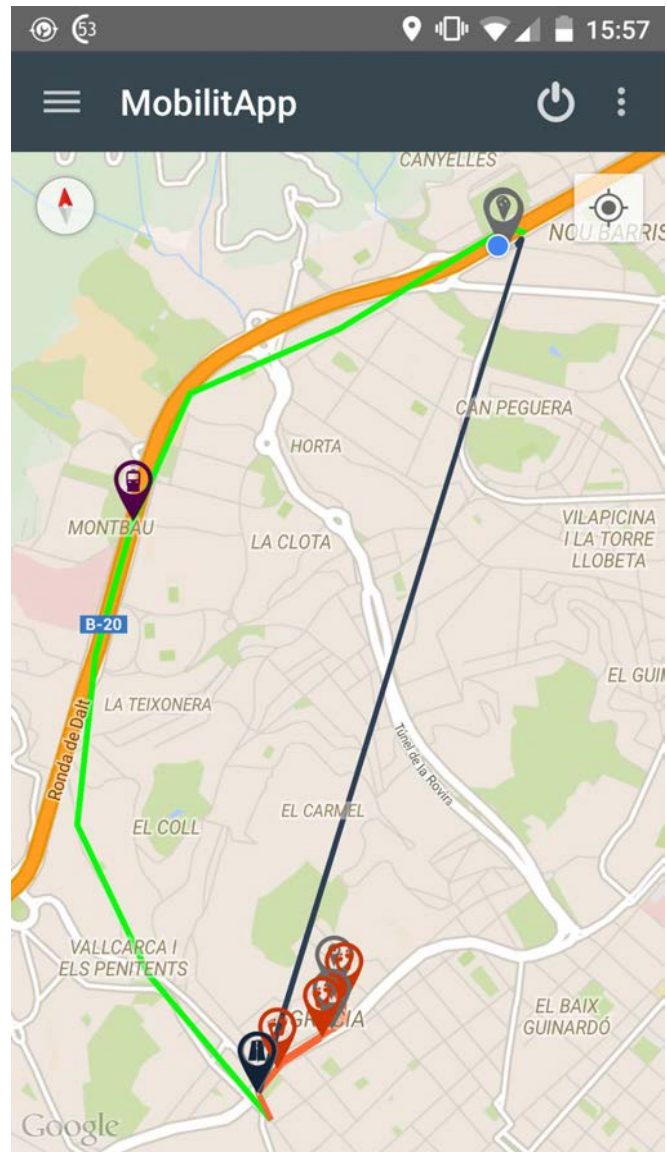

The Metro ride has been detected perfectly the first time but not in the return trip. Metro detection is triggered when consecutive bad locations are registered (this is the normal behaviour in the Metro) but sometimes location gets stuck and the same location (i.e. the last good one location) is received until we leave the metro. In this situation Metro recognition is not triggered and then we only have 2 different locations and consequently a straight line is displayed on the map.

This misbehaviour will be corrected in future versions of MobilitApp with an improved algorithm.

### A.3.    Journey 3

**Information:** Home-UPC journey (round trip) by bus (Line 60).

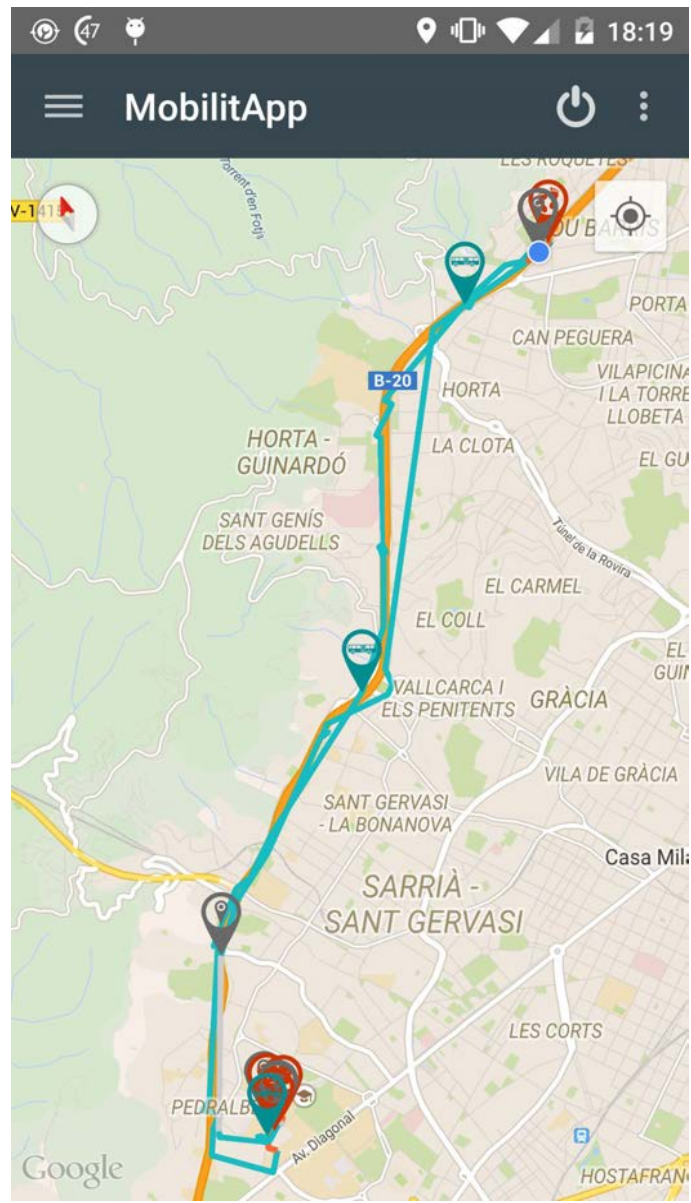

In both cases bus recognition was successful (one more accurate than the other). There is a section with still activity in the return trip because the bus was still a long time.

It is important to know that all road journeys require the GPS because there is no Wi-Fi nearby. To avoid huge battery drain, GPS is enabled every 5 seconds for a maximum of 4 minutes.

### A.4.    Journey 4

**Information:** Walk journey without using any transport.

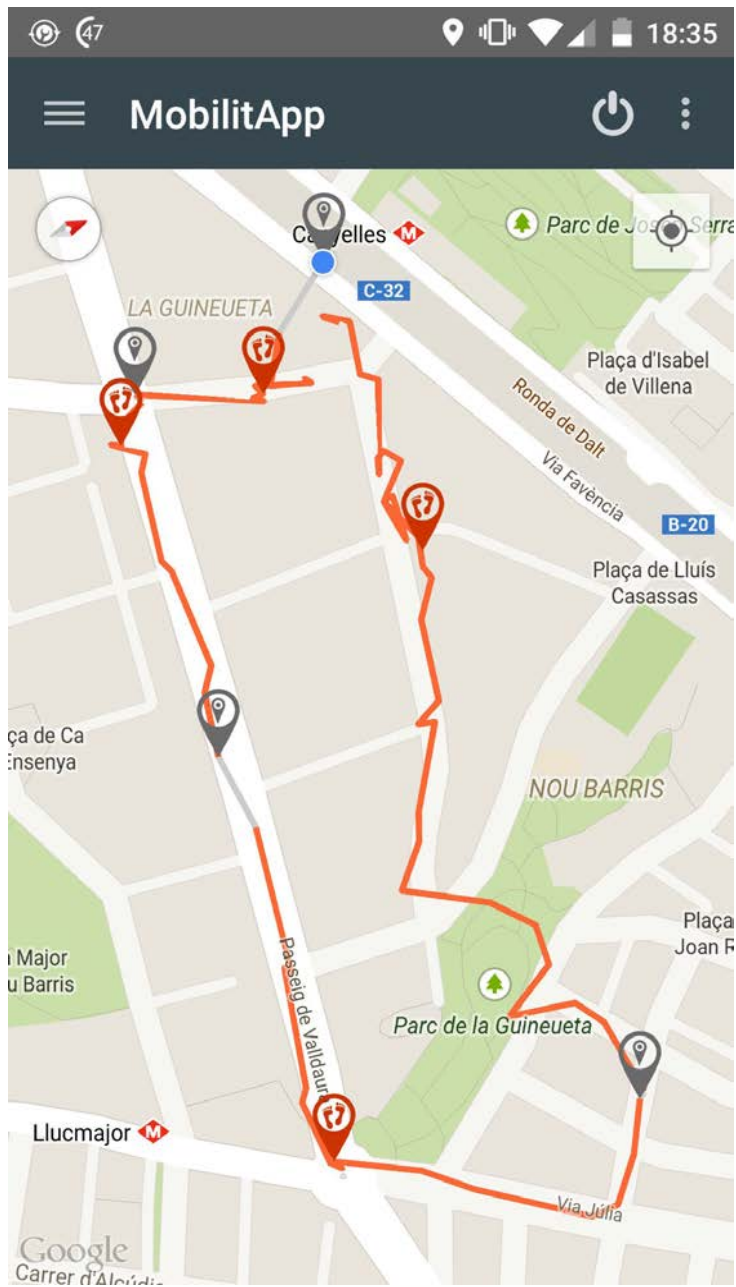

GPS was not required in this journey; Wi-Fi is sufficient for this kind of journeys. Wi-Fi positioning has a block level accuracy of 100 meters; this is the reason of some "peaks" on the map.

## A.5.    Journey 5

**Information:** Journey by car.

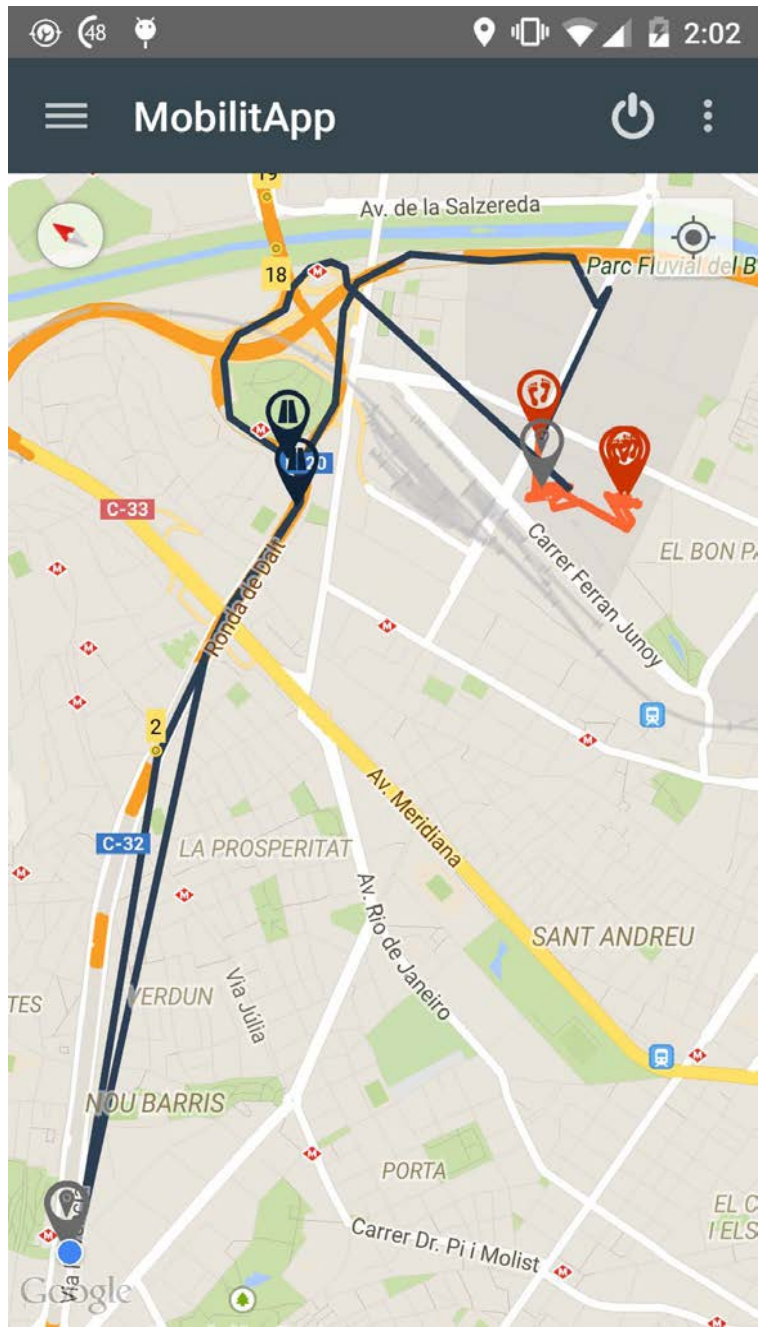

Conversely, GPS was required during this journey by car. We only notice good locations in the middle of the journey because GPS takes a little time to be 100% functional and not always good locations are received even in favourable conditions.

# Annex B

# MobilitApp User Guide

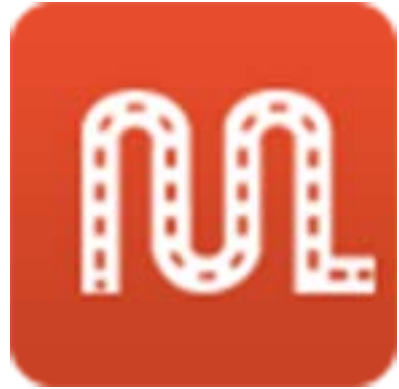

# MobilitApp Android App

# User Guide

## B.1. Installation

First we need to enable Unknown Sources (in Spanish, *orígenes desconocidos).*
To enable Unknown Sources go to Settings > Security and check the box next to" Unknown Sources ". A dialogue box may pop-up asking you to confirm the action, just tap OK to confirm.

MobilitApp can be downloaded in APK (Android Application package) format in the following URL (Simulators and Code section):

<div align="center">http://sertel.upc.edu/~maguilar/indexold.html</div>

Here, we can also download the final degree project where all related with MobilitApp is explained.

Once you have the APK file downloaded on your laptop or PC, connect your Android device to the computer and copy (**drag and drop**) the APK file to the external/internal storage on your device (for example in Downloads, Music folder, etc.).
-   In Mac OS, you would need an application named **Android File Transfer,** that can be found in: **https://www.android.com/filetransfer/**
-   In Windows, if you connect your Android device to the USB port, once the drivers have been installed (automatic), you could browse your device like it was a pendrive.

**IMPORTANT** -> If your Android device does not have a file explorer or similar, you need to go to the Play Store (in your device) and search and download **ES File Manager/ ES Explorador de Archivos** (in spanish).

Open File Manager and navigate to the directory where you have copied or downloaded your APK file. Mostly, it's stored in the "Downloads" folder on your Android device (for instance -> **/storage/emulated/0/Download** is my Download Folder). If your device has an SD card maybe you need to navigate to **/storage/sdcard0**. Once in the directory, tap the APK icon and an installation dialogue box will pop-up. Hit the install button to install the APK on your Android.

### B.2. Log in and preferences

After installation, click on the MobilitApp icon to run the app. Now, you will see the login:

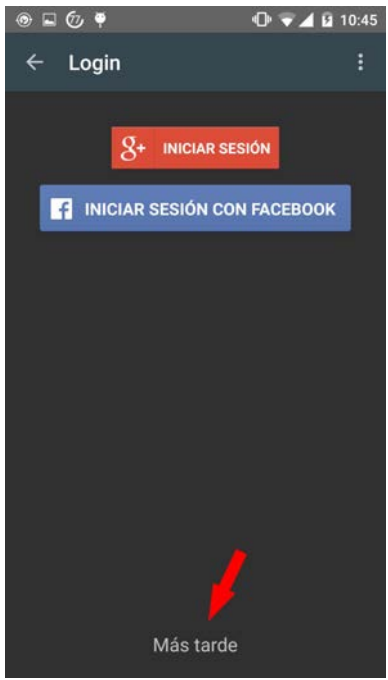

• You can log in with your Gmail account or with Facebook.

• Also you can press Later (*Más tarde* in Spanish) to log in later.

Once the login is done or Later is pressed, the preferences screen will pop up.

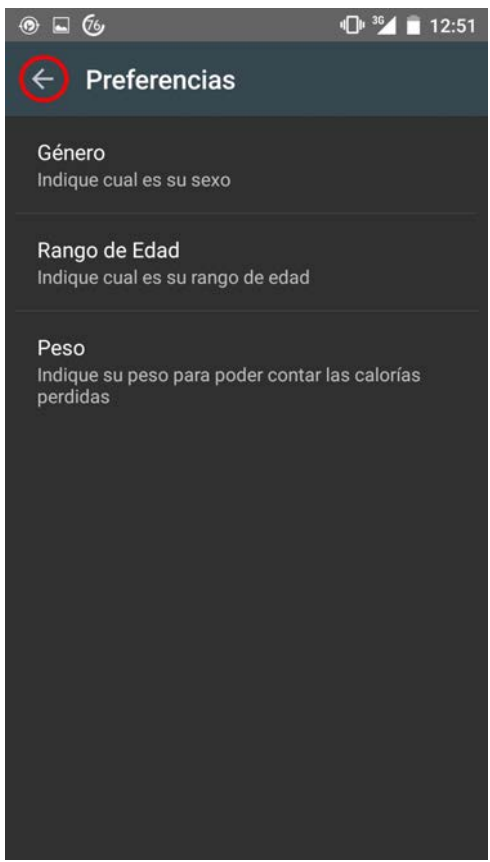

• Here we can select our gender, age range and set our weight.

• Weight is important because is needed to compute/estimate the number of calories that you burn (on foot activity).

• To continue, we need to press the top left arrow (marked with a red circle in the image)

## B.3. General functions

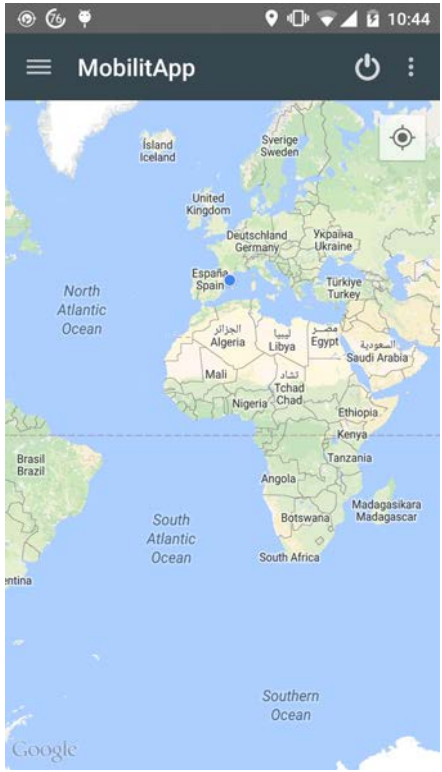

• Initial screen with the toolbar, the map and the icon to zoom the map.

• The toolbar is used to give functionality to MobilitApp

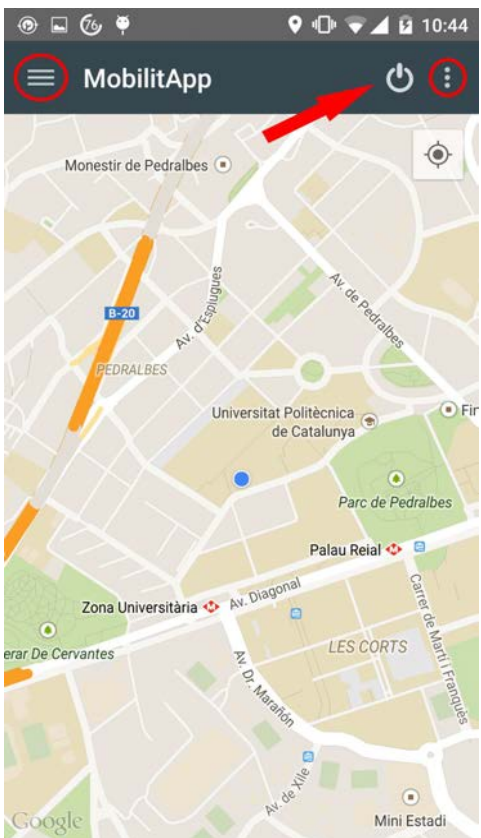

• That toolbar consists in:

• Top-left icon: The navigation menu appears when pressed (showed in the following screenshot).

• **IMPORTANT: Shutdown icon** used to completely finish the Application. By clicking that icon, the services used to track the position and guess your activity finish.

• Top-right icon: Used to show again the preferences screen in order to modify any wrong values.

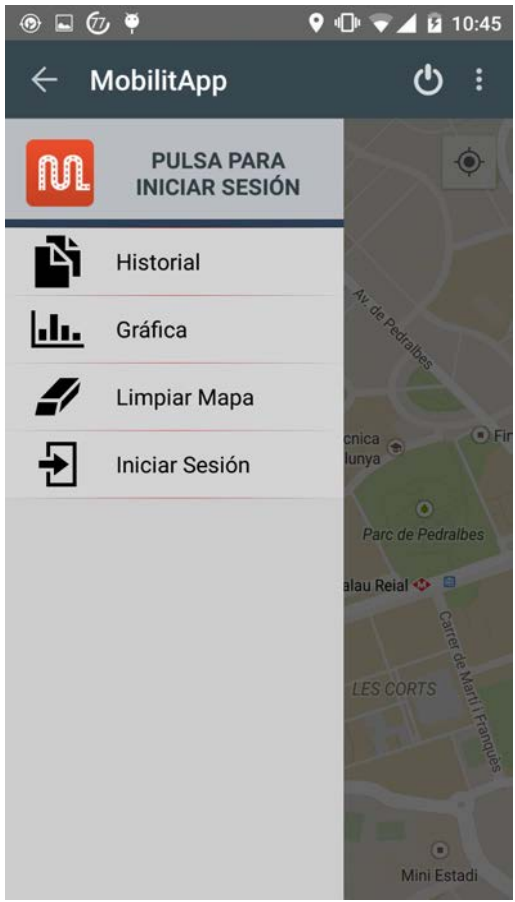

• This navigation menu is showed when you have not logged in.

• You can log in by pressing the text that appears to the right of the logo or in Log in (*Iniciar Sesión* in Spanish).

• **History:** Go to History screen.

• **Plots:** Go to Plot screen.

• **Clean Map:** cleans the map when markers and lines are showed.

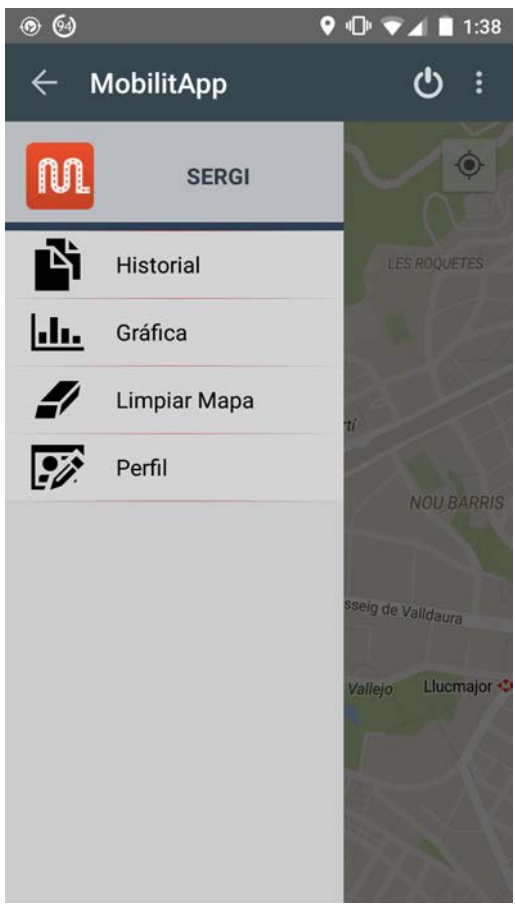

1. Appears when you are logged in.

2. The main difference is that we go to the profile screen when Profile (*Perfil* in Spanish) or your name is pressed.

### B.3.1. History

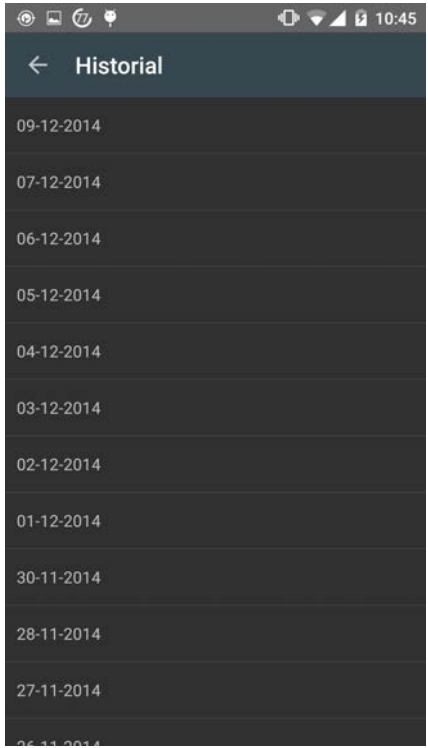

• By pressing the different days you will see in the map your approximate route in that day with the different activities (on foot, still, by bus...).

• These files are locally stored in your device so, in order to delete this files you can longclick/longpress and a button will appear. Note that you can only delete files created (at least) **3 days ago.**

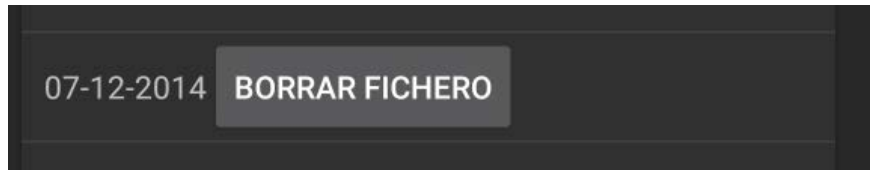

Let's press one day and something like this will be showed:

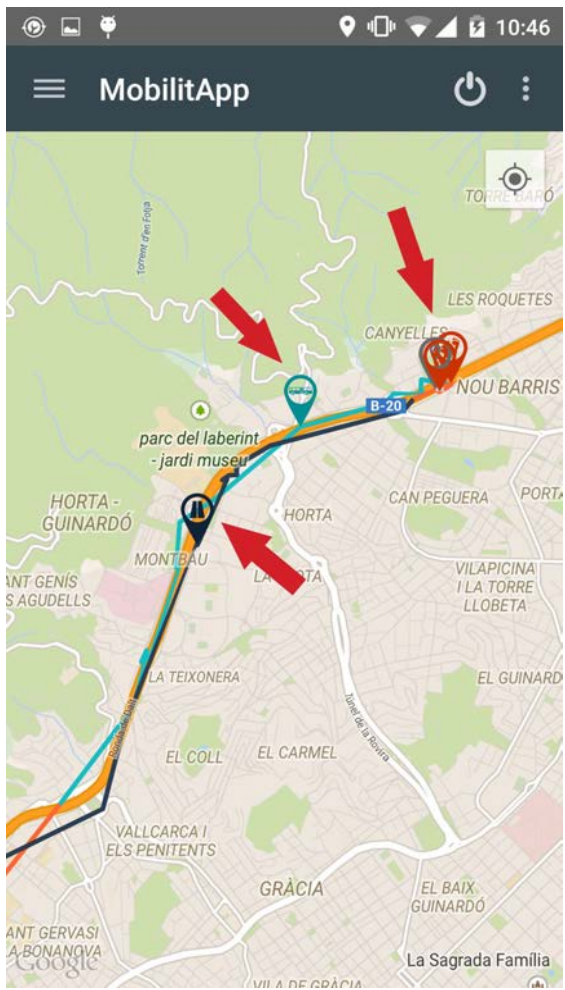

• For each activity you have done a marker will be displayed over the map and also, a line with your approximate route.

• If you want to see the information of those routes you only need to press the marker.

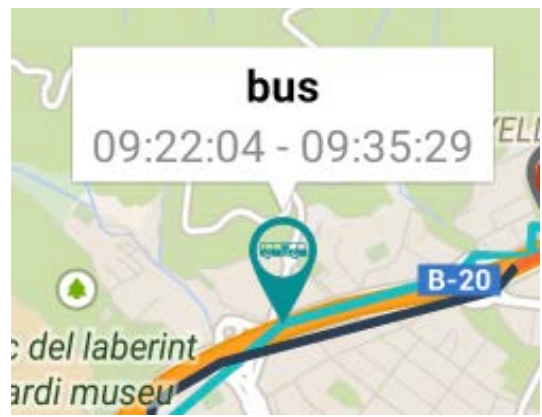

If we want more detailed information we can press that pop up window and we will see: speed, distance, duration, co2 saved or calories burned.

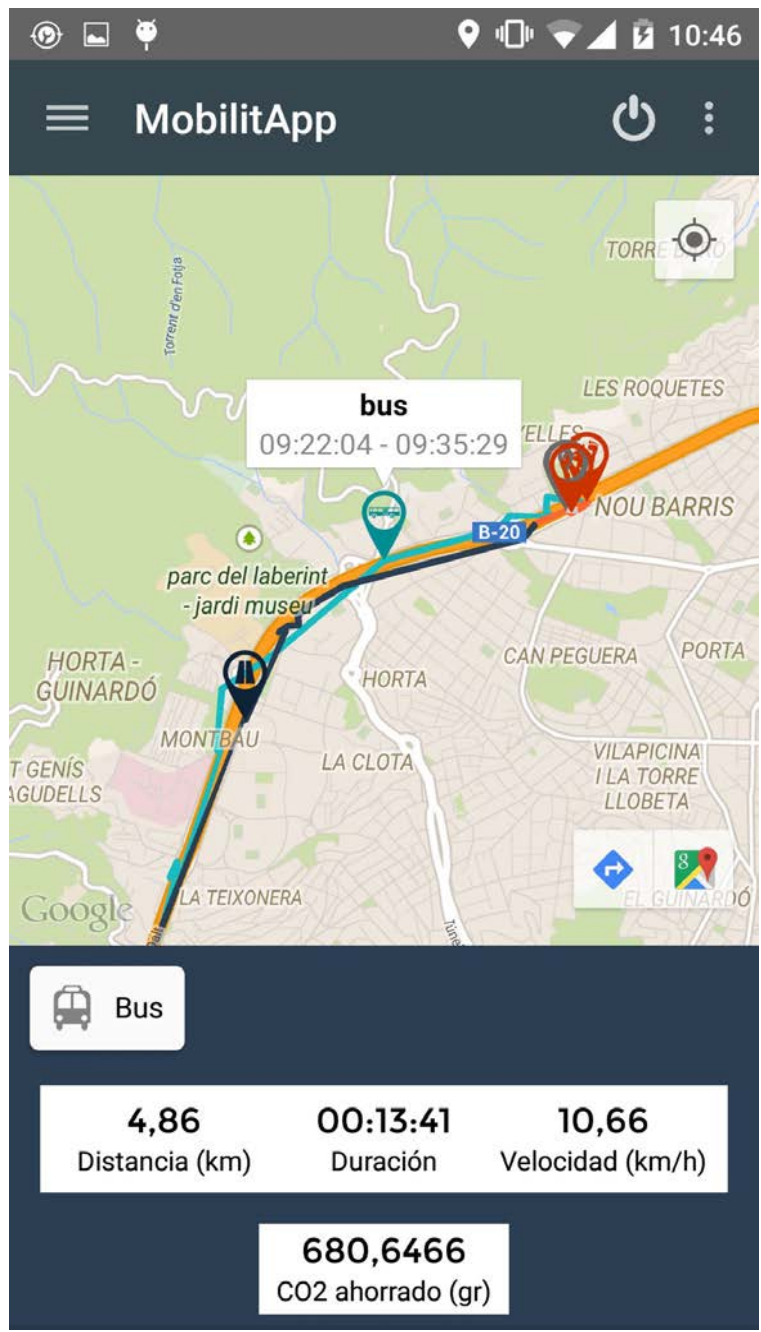

To return to full map we must simply press some part of the map.

Note: These two new icons are not part of MobilitApp, if you press them you will go to Google Maps application.

### B.3.2. Plot

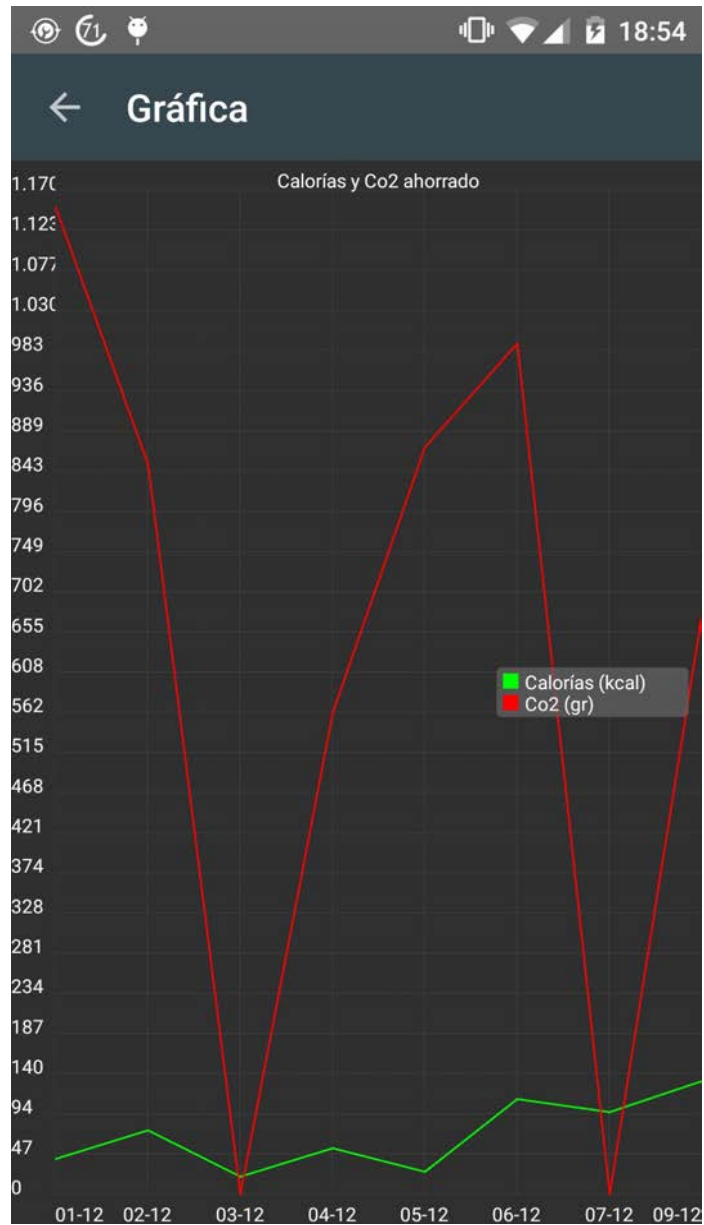

Plot to show calories burned (in kcal) and co2 saved (in gr.).

### B.3.3. Profile

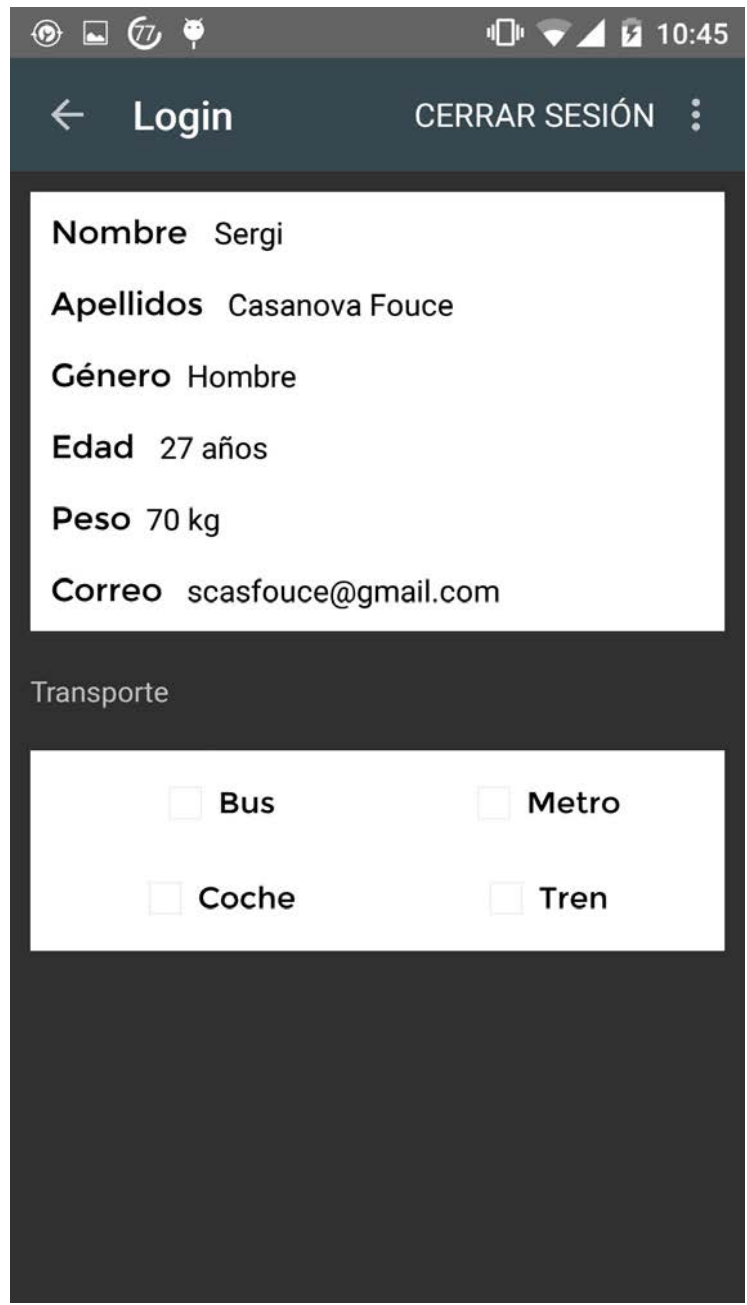

Profile screen shows user information based on your login/preferences. In future versions of MobilitApp that information may be changed (only in MobilitApp scope).

We can close Facebook or Gmail session by pressing the Log Out button (*Cerrar sesión* in spanish).

Note: The transport part (below the profile) is non-functional so, pressing that will not affect in the application.